\documentclass{fundam}

\usepackage{latexsym,amsmath,amssymb}
\usepackage{graphicx}
\usepackage[utf8]{inputenc}
\usepackage{url}

\newtheorem{defin}{Definition}
\newtheorem{op_pb}{Open problem}

\newcommand{\F}{\mathbb{F}}
\newcommand{\x}{\mbox{$\bf x$}}
\newcommand{\y}{\mbox{$\bf y$}}
\newcommand{\bc}{\mbox{$\bf c$}}
\newcommand{\bu}{\mbox{$\bf u$}}

\newcommand{\Idr}[2]{Id_{#2}(#1)}
\newcommand{\Idrl}[3]{Id_{(#2, \leq #3)}(#1)}

\usepackage{hyperref}

\begin{document}


\setcounter{page}{165}
\publyear{24}
\papernumber{2178}
\volume{191}
\issue{3-4}

\finalVersionForARXIV


\title{On Iiro Honkala's Contributions to Identifying Codes}

\author{Olivier Hudry\thanks{Address for correspondence: Département Informatique et Réseaux de Télécom Paris,
 France.}
\\
	Télécom Paris, France\\
	olivier.hudry@telecom-paris.fr
	\and
	Junnila Ville\\
	University of Turku, Finland\\
	viljun@utu.fi
   \and
 Antoine Lobstein\\
 CNRS, Universit\'e Paris-Saclay, France\\
 lobstein@lri.fr}

\maketitle

\runninghead{O. Hudry  et all.}{On Iiro Honkala's Contributions to Identifying Codes}

\vspace*{-6mm}
\begin{abstract}
A set $C$ of vertices in a graph $G=(V,E)$ is an identifying code if it is dominating and any two vertices of $V$ are dominated by distinct sets of codewords. This paper presents a survey of Iiro Honkala's contributions to the study of identifying codes with respect to several aspects: complexity of computing an identifying code, combinatorics in binary Hamming spaces, infinite grids, relationships between identifying codes and usual parameters in graphs, structural properties of graphs admitting identifying codes, and number of optimal identifying codes.

 \end{abstract} 

\section{Introduction}
\label{IdLD6Intro}

The graphs $G=(V,E)$ ($V$ is the vertex set of $G$ and $E$ its edge set) that we shall consider in this article are undirected and, unless otherwise stated, connected. A {\it code} is any subset of $V$, whose elements are called {\it codewords}. For any integer $r \geqslant 1$, the \textit{ball of radius }$r$, or the $r$-\textit{ball}, of a vertex $v\in V$ is the set ${B_r(v)=\{u \in V: 0 \leqslant d(u,v) \leqslant r\}}$, where the considered distance $d$ is the one provided by the length of a shortest path between $u$ and $v$ (for $r = 1$, $B_1(v)$ is the same as the usual {\it closed neighbourhood} $N[v]$). Two distinct vertices $v_1$ and $v_2$ are said to be $r$-{\it twins} if they share the same $r$-ball: $B_{r}(v_1)=B_{r}(v_2)$.

\begin{defin}
For any code $C \subseteq V$ and any vertex $v \in V$, the $r$-identifying set $I_r(C;v)$ of $v$ with respect to $C$ is defined by $I_r(C;v) = B_r(v) \cap C$. Furthermore, for any subset $S$ of $V$, we set: ${I_r(C;S) = \bigcup_{v \in S} I_r(C;v)}$.
\end {defin}

In other words, $I_r(C;v)$ is the set of codewords which are at distance at most $r$ from $v$. When there is no ambiguity, we may write $I_r(v)$ and $I_r(S)$ instead of $I_r(C;v)$ and $I_r(C;S)$ respectively.

Two vertices within distance~$r$ from each other are indifferently said to $r$-{\it dominate} or to $r$-{\it cover} each other. A vertex $v$ is $r$-\textit{dominated} or $r$-\textit{covered} {\it by a code} $C$ if it is $r$-dominated by at least one element of $C$, i.e., if its $r$-identifying set is not empty: $I_r(C;v) \neq \emptyset$. If every vertex is $r$-dominated by $C$, then $C$ is said to be $r$\textit{-dominating} or $r$\textit{-covering}. Two vertices $u$ and $v$ are $r$-\textit{separated} by a vertex $w$ if we have $w \in B_r(u)$ and $w \notin B_r(v)$ or if we have $w \in B_r(v)$ and $w \notin B_r(u)$ (note that $w$ can be equal to $u$ or $v$). Two vertices $u$ and $v$ are $r$-separated \textit{by a code} $C$ if they are $r$-separated by at least one element of $C$: $I_r(C;u) \neq I_r(C;v)$. If every distinct vertices $u$ and $v$ are $r$-separated, then $C$ is said to be $r$\textit{-separating}.	

\medskip
The conjunction of the $r$-domination and of the $r$-separation properties leads to the definition of an $r$-identifying code:

\begin{defin}
An $r$-{\it identifying code}~$C\subseteq V$, or simply an $r$-IdC, is an $r$-dominating and $r$-separating code. In other words, $C$ fulfills the following two properties:
\begin{itemize}
\itemsep=0.9pt
\item $\forall \, v \in V: I_r(C;v) \neq \emptyset$;
\item $\forall \, v_1 \in V, \, \forall \, v_2 \in V \text{ with } v_1 \neq v_2: \; I_r(C;v_1) \neq I_r(C;v_2).$
\end{itemize}
\end{defin}

It is easy to check that a graph~$G$ admits an $r$-IdC if and only if it has no $r$-twins (then, $V$ itself is an $r$-IdC), i.e.:
\[ 
  \forall \, v_1\in V, \; \forall \, v_2 \in V \text{ with } v_1 \neq v_2: \; B_{r}(v_1) \neq B_{r}(v_2).
\] 

For any $r$-twin-free graph $G$, we denote by $Id_{r}(G)$ the smallest possible cardinality of an $r$-IdC and we call this the $r$-{\it identification number} of~$G$, or simply the $r$-Id number of $G$. 
An $r$-IdC with the smallest possible size, i.e. whose cardinality is equal to $Id_{r}(G)$, is said to be {\it optimal}.

\begin{example}
Consider the graphs provided by Figure~\ref{fig6example1}.\vspace{1mm}
\begin{figure} [h!]
\begin{center}
\includegraphics*[scale=0.9]{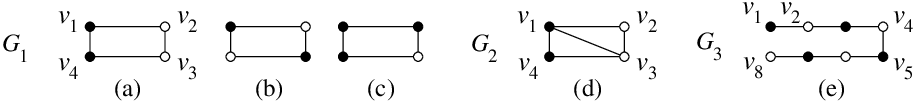}\vspace*{-6mm}
\end{center}
\caption{Different graphs and codes. Black vertices represent codewords.}
\label{fig6example1}
\end{figure}

\begin{itemize}
\item For $G_1$: clearly, $G_1$ requires at least two vertices in order to 1-cover all the vertices. But: (a)~$\{v_1,v_4\}$ is not 1-identifying since $v_1$ and $v_4$ are not 1-separated; (b)~$\{v_1,v_3\}$ is not 1-identifying since $v_2$ and $v_4$ are not 1-separated; so any 1-IdC contains at least three vertices; (c)~$\{v_1,v_2,v_4\}$ is 1-identifying since all the vertices are 1-covered and any two distinct vertices are 1-separated. Hence $Id_1(G_1)=3$. For $r > 1$, all the vertices of $G_1$ are $r$-twins.


\item For $G_2$:~$v_1$ and $v_4$ are not 1-separated by the code displayed in (d). Moreover, $v_1$ and $v_3$ are 1-twins: therefore they cannot be 1-separated. For $r > 1$, all the vertices of $G_2$ are $r$-twins. Thus, $G_2$ admits no $r$-IdC for any $r \geqslant 1$.
\item For $G_3$: the code $\{v_1, v_3, v_5, v_7\}$ displayed in (e) is not 1-identifying since $v_7$ and $v_8$ are not 1-separated; but $\{v_1, v_3, v_5, v_6, v_7\}$ is 1-identifying. In contrast to the case $r=1$, $v_7$ and $v_8$ are now 2-separated by $\{v_1, v_3, v_5, v_7\}$ (more precisely, by $v_5$), but not $v_6$ and $v_7$. For $r \geqslant 4$, $v_4$ and $v_5$ are $r$-twins.
\end{itemize}
\end{example}

A usual illustration of identifying codes comes from fault diagnosis in multiprocessor networks. Such a network can be represented as a graph whose vertices are processors and edges are links between processors. Assume that a processor is malfunctioning; we wish to locate the faulty processor. For this purpose, some processors (which will constitute the $r$-IdC) will be selected and assigned the task of testing the vertices at distance at most $r$. Whenever a selected processor (i.e., a codeword) detects a fault, it sends a binary message saying that one element in its $r$-ball is malfunctioning. We require that we can unambiguously identify the location of the malfunctioning processor, based only on the information which ones of the codewords sent a message; in this case, an identifying code is what we need.

The term ``identifying code'' is used in the 1998 paper~\cite{karp98a} by M.~G.~Karpovsky, K.~Chakrabarty and I.~B.~Levitin, which certainly marks the starting point for the blossoming of works on this topic, but the concept is already contained in~\cite{rao93} by N.~S.~V.~Rao in 1993 (for references on identifying codes, see the ongoing bibliography at~\cite{http}). 

The aim of this paper, in this issue devoted to Iiro Honkala, is to survey some of Iiro's contributions to the study of identifying codes. For this reason, when quoting a reference, we add a star (for instance \cite{auge09b}*) for denoting that this contribution is authored or co-authored by Iiro. For the same reason, some aspects of identifying codes are not covered here (for a wider survey on IdCs, see, e.g., \cite{lobsteinLocatingdominationIdentification2020}).

\medskip
We consider also a generalization of $r$-IdCs, which occurs when more than one vertex should be identified~\cite{karp98a}: for a given integer $\ell$, an ($r,\leqslant \ell$)-{\it identifying code}, or simply ($r,\leqslant \ell$)-{\it IdC}, in a graph $G=(V,E)$ is an $r$-dominating code $C$ such that for all $X\subseteq V$, $Y \subseteq V$ with $X \neq Y$, $0 < |X| \leqslant \ell$, $0 < |Y|\leqslant \ell$, we have:
$$\bigcup _{u\in X} I_{r}(C;u) \neq \bigcup _{v\in Y} I_{r}(C;v).$$
As for $r$-IdCs, we say that an ($r,\leqslant \ell$)-IdC is \textit{optimal} if its size is minimum. Then $Id_{(r,\leqslant \ell)}(G)$ denotes this minimum size.

\medskip
This generalization, and other similar ones, have been studied mostly in the $n$-cube and in the grids; we refer to, e.g., \cite{exoo10}, \cite{exoo08d},~\cite{fouc13}, \cite{grav05}, \cite{honk04d}*, \cite{honk04a}*, \cite{karp98a}, \cite{laih02a}, \cite{pelt10}.

The following theorem provides a general property fulfilled by $r$-twin-free graphs. 

\begin{theorem}[\cite{Auger_2008}] \label{6easyId61} Let $r\geqslant 1$ be any integer. Any $r$-twin-free graph $G$ with at least one edge admits $P_{2r+1}$, the path on 2r + 1 vertices, as an induced subgraph. As a consequence, $G$ has order $n\geqslant 2r+1$ and the only $r$-twin-free graph with order $n=2r+1$ and with at least one edge is the path $P_{2r+1}$.
\end{theorem}

Theorem~\ref{lemmeNath6} gathers a lower bound from~\cite{karp98a} 
and an upper bound from~\cite{bert01} for $Id_{r}(G)$, both sharp (see also~\cite{char04b} and~\cite{grav07}).

\begin{theorem} \label{lemmeNath6} For any $r\geqslant 1$ and any connected $r$-twin-free graph~$G=(V,E)$ of order~${n \geqslant 2r + 1}$:
$$\lceil \log _2 (n+1)\rceil \leqslant Id_{r}(G) \leqslant n-1.$$ \end{theorem}

In what follows, the structure of the paper is explained. In Section~\ref{IdLD6Complex}, the computational complexity of determining an $r$-identifying code of a given size. Then, in Section~\ref{Sec_Hamming}, we proceed with various results concerning binary Hamming spaces. In Section~\ref{IdLD6Specif4}, the results for infinite grids --- the square, triangular, king and hexagonal grids --- are summarized. As the existence of an identifying code requires the underlying graph to be twin-free, the structural properties of such graphs are studied in Section~\ref{IdLD6SI1}. Although the question of determining an optimal identifying code in a graph is already rather difficult, its natural extension of calculating the number of optimal codes has also been previously studied as examined in Section~\ref{IdLD6NumbSol}. Finally, in Section~\ref{IdLD6Conc}, some variants and closely related concepts of identifying codes are briefly discussed.

\section{Complexity for general graphs} \label{IdLD6Complex}
The decision problem associated with the computation of $Id_{r}(G)$ is NP-complete for any fixed $r \geqslant 1$, as stated by Theorem~\ref{6ourcompl6}.

\begin{theorem} \label{6ourcompl6}
  {\rm(\cite{cohe01a}* for $r=1$, \cite{char03b} for $r>1$}) For any fixed $r \geqslant 1$, the following problem is NP-complete:

\medskip
Name: $r$-IdC

Instance: A graph $G=(V,E)$ and an integer $k \leqslant |V|$.

Question: Does $G$ admit an $r$-IdC of size at most~$k$?
\end{theorem}

\begin{figure}[ht!]
\vspace*{-2mm}
\begin{center}
\includegraphics*[scale=0.7]{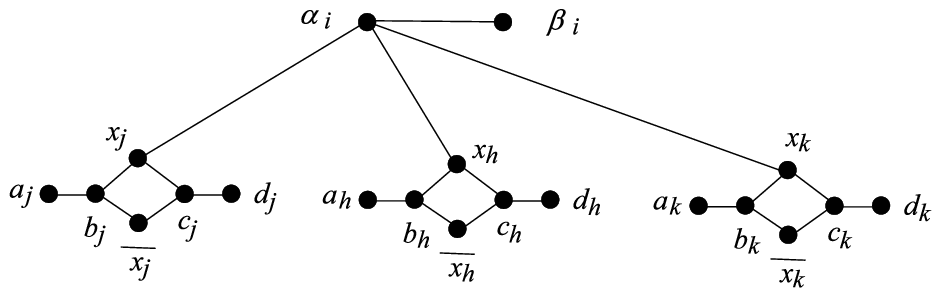}
\end{center}\vspace*{-8mm}
\caption{The subgraph induced by a clause $C_i = x_j \vee x_h \vee x_k$ for the transformation from 3-SAT to $1$-IdC.}
\label{Fig_Compl}\vspace*{-3mm}
\end{figure}

\noindent The proof of Theorem~\ref{6ourcompl6} for $r=1$ is based on a polynomial transformation from the NP-complete problem 3-SAT. Let $(U; C_1, C_2, ..., C_m)$ be any instance of 3-SAT, where $U$ denotes the set of Boolean variables and where $C_1, C_2, ..., C_m$ are the clauses, each clause containing three literals. The transformation uses two kinds of components. The first one, associated with each Boolean variable $x_j$ ($1 \leqslant j \leqslant |U|$), contains six vertices named $x_j$, $\overline{x_j}$, $a_j$, $b_j$, $c_j$ and $d_j$ and the six edges $\{a_j, b_j\}$, $\{b_j, x_j\}$, $\{b_j, \overline{x_j}\}$, $\{c_j, d_j\}$, $\{c_j, x_j\}$, $\{c_j, \overline{x_j}\}$. The second one, associated with each clause $C_i$ ($1 \leqslant i \leqslant m$), contains two vertices named $\alpha_i$ and $\beta_i$ and the edge $\{\alpha_i, \beta_i\}$. Extra edges reflect the composition of any clause: if $C_i$ contains the three literals $x_j$, $x_h$ and $x_k$, we add the edges $\{\alpha_i, x_j\}$, $\{\alpha_i, x_h\}$, $\{\alpha_i, x_k\}$. Figure~\ref{Fig_Compl} illustrates this transformation by specifying a part of the graph thus obtained.

\medskip
By setting $k = 3|U| + m$, it is proved in \cite{cohe01a}* that the instance of 3-SAT is satisfiable if and only if there exists a 1-IdC of size at most $k$ in the constructed graph (in fact the size is then exactly $k$ and such a code can be provided by all the vertices $\alpha_i$ with $1 \leqslant i \leqslant m$, the vertices $x_j$ or $\overline{x_j}$ according to the Boolean value taken by the variable associated with $x_j$ and $\overline{x_j}$ for $1 \leqslant j \leqslant |U|$, and all the vertices $b_j$ and $c_j$ for $1 \leqslant j \leqslant |U|$). This construction is generalized in~\cite{char03b} to any integer~$r$.

The NP-completeness of the problem $r$-IdC involves that, given $G$ and $r$, determining $Id_r(G)$ or finding an optimal $r$-IdC of $G$ are NP-hard problems (for other results related to complexity issues, see~\cite{Auger_et_al_2010}, \cite{fouc12}, \cite{fouc15}, \cite{Hud_Lob_2016} and \cite{Hud_Lob_2019}).

Specific families of graphs have been studied, including the binary Hamming spaces considered in the next section. For complexity results dealing with identification in these graphs, see \cite{honk02b}* and~\cite{honk02c}* and, below, Subsection~\ref{Compl_Hamming}.

\section{Binary Hamming spaces} \label{Sec_Hamming}


For the whole section, let $n$ be a positive integer. The \emph{binary Hamming space} $\F^n$ is the $n$-fold Cartesian product of the binary field $\F= \{ 0, 1 \}$. The $2^n$ elements of $\F^n$, i.e. vectors with $n$ binary coordinates, are called \emph{words}. A subset of $\F^n$ is called a \textit{code of length} $n$. Let $\x$ and $\y$ be words belonging to $\F^n$; the \emph{Hamming distance} $d(\x, \y)$ between words $\x$ and $\y$ is the number of coordinate places in which they differ. The \emph{Hamming ball} of radius $r$ centred at $\x$ is denoted by $B_r(\x)$ and consists of the words that are within Hamming distance $r$ from $\x$.

The Hamming space may also be seen as a graph (observe then that $n$ does not denote the number of vertices of the graph, but still the length of the Hamming space): $\F^n$ is the vertex set; if $\x$ and $\y$ are words belonging to $\F^n$ (i.e. two vertices of the associated graph), there is an edge between $\x$ and $\y$ if $\x$ and $\y$ differ by one coordinate, i.e. if their Hamming distance is equal to 1. As this graph is completely defined by $\F^n$, we will denote it by $\F^n$, omitting its edge set. In $\F^n$, the shortest path distance is the same as the Hamming distance and, thus, the Hamming ball of radius $r$ is the same as the $r$-ball defined in Section~\ref{IdLD6Intro}. Hence the same notation is used in this section for the distance $d$ and the ball $B_r$ as in Section~\ref{IdLD6Intro}.

The set of non-zero coordinates of the word $\x$ is called the \emph{support} of $\x$ and is denoted by $supp(\x)$. The \emph{weight} of $\x$ is the cardinality of the support of $\x$ and is denoted by $w(\x)$.

\medskip
The size of a Hamming ball of radius $r$ in $\F^n$ does not depend on the choice of the centre and it is denoted by $V(n,r)$. Furthermore, we have:
\[
V(n,r) = \sum_{i=0}^r \binom{n}{i} \textrm{.}
\]
\eject
Assuming $\x = (a_1, a_2, \ldots, a_n) = a_1a_2\cdots a_n$ and $\y = (b_1, b_2, \ldots, b_m) = b_1 b_2 \cdots b_m$, we can define the \emph{concatenation} $(\x, \y)$ of the words $\x$ and $\y$ as follows:
\[
(\x, \y) = (a_1, a_2, \ldots, a_n, b_1, b_2, \ldots, b_m) = a_1 a_2 \cdots a_n b_1 b_2 \cdots b_m \textrm{.}
\]
The \emph{sum} of the words $\x = (a_1, a_2, \ldots, a_n)$ and $\y = (b_1, b_2, \ldots, b_n)$ is defined as
\[
\x + \y = (a_1 + b_1, a_2 + b_2, \ldots, a_n + b_n)
\]
where the sum $a_i + b_i$ ($1 \leqslant i \leqslant n$) is performed modulo 2.

\medskip
If $C_1 \subseteq \F^n$ and $C_2 \subseteq \F^m$ are codes, then their \emph{direct sum}
\[
C_1 \oplus C_2 = \{ (\x,\y) \mid \x \in \F^n, \y \in \F^m\} \subseteq \F^{n+m}
\]
is defined as a code in $\F^{n+m}$. The following well-known lemma, which considers the intersections of balls of radius $1$ in $\F^n$, is often quite useful.
\begin{lemma} \label{Lemma_intersection_balls}
	For $\x \in \F^n$ and $\y \in \F^n$:
	\[
	|B_1(\x) \cap B_1(\y)| = \begin{cases}
		n+1 \text, & \text{if } d(\x,\y) = 0. \\
		2 \text, & \text{if } 1 \leqslant d(\x,\y) \leqslant 2. \\
		0,  & \text{if } 2 < d(\x,\y).
	\end{cases}
	\]
Furthermore, the intersection of three distinct Hamming balls of radius $1$ contains at most one word.
\end{lemma}

Note that $(r, \leqslant \ell)$-IdCs do not exist for all lengths $n$. For $\ell \geqslant 1$, the existence of an $(r, \leqslant \ell)$-IdC is discussed in the following theorem.

\setcounter{footnote}{0} 
\begin{theorem}[\cite{honkalaCodesIdentifyingSets2001}*]
	Let $r(n,K)$ denote the smallest radius $r$ such that there exists an $r$-covering code in $\F^n$ with cardinality $K$. If $1 \leqslant \ell < 2^n$ and $r \geqslant r(n,\ell)$, then there exists no $(r, \leqslant \ell)$-IdC in~$\F^n$.
\end{theorem}

Extensive coverage of the values $r(n,K)$ can be found in~\cite{cohenCoveringCodes1997}*. For example, if $\ell = 1$, then an $r$-IdC in $\F^n$ exists if and only if $n \geqslant r+1$.
Based on these results, we also obtain that there exists no $(1, \leqslant 2)$-IdC of length smaller than $4$ and no $(2, \leqslant 2)$-IdC of length smaller than $6$.

\medskip

In the rest of this section, we first focus on complexity issues for Hamming spaces. Then, in Subsection~\ref{Subsec_Hamming_lower_bounds}, various lower bounds on $r$-IdCs are considered. In Subsection~\ref{Subsec_Hamming_upper_bounds}, we proceed with the constructions of $r$-IdCs leading to the best known upper bounds. In Subsection~\ref{Subsec_Hamming_conjectures}, we discuss some natural conjectures concerning identification in binary Hamming spaces and consider the progress made towards them. Finally, in Subsections~\ref{Subsec_larger_ell} and~\ref{other_variants}, we study codes identifying sets of vertices, and briefly discuss some other variants of (usual) $(r, \leqslant 1)$-IdCs.

\subsection{Complexity issues for binary Hamming spaces} \label{Compl_Hamming}
As stated in Section~\ref{IdLD6Complex}, the problem of deciding whether there exists an $r$-IdC in a general graph with at most $k$ codewords is NP-complete. The problem remains algorithmically difficult when we restrict ourselves to binary Hamming spaces. More precisely, the following theorem is proved in~\cite{honk02c}* thanks to a polynomial transformation from 3-SAT.

\begin{theorem} 
  [\cite{honk02c}*] The following problem is NP-complete:

\medskip
Name: Not $r$-IdC

Instance: An integer $n$, a list of binary codewords of length $n$, forming a code $C$; an integer $r$.

Question: Are there distinct words $\x$ and $\y$ belonging to $\mathbb{F}^n$ with $B_r(\x) \cap C = \emptyset$ or with ${B_r(\x) \cap C = B_r(\y) \cap C}$?
\end{theorem}

Observe that, for the binary Hamming spaces, it is not necessary to encode the graph by its adjacency matrix, as usually done for an arbitrary graph. Indeed, the specification of the length $n$ suffices to characterize $\mathbb{F}^n$. In fact, even the problem of deciding whether a given code is $r$-identifying in $\F^n$ is co-NP-complete.

\subsection{Lower bounds} \label{Subsec_Hamming_lower_bounds}

In this subsection, we discuss various lower bounds concerning $r$-IdCs in $\F^n$. The first one of these lower bounds presented in the following theorem is based on the simple fact that if $C \subseteq \F^n$ is an $r$-IdC, then there are at most $|C|$ words $r$-covered by exactly one codeword and all the other words of $\F^n$ have to be $r$-covered by at least two codewords.

\begin{theorem}[\cite{karp98a}] \label{Thm_2-fold_lower_bound}
We have $\Idr{\F^n}{r} \geqslant \dfrac{2^{n+1}}{V(n,r)+1}$.
\end{theorem}

This lower bound can be further refined by observing that if $C \subseteq \F^n$ is an $r$-IdC, then the number of words $r$-covered by exactly $k$ codewords is restricted by the binomial coefficient $\binom{|C|}{k}$. This result has been originally presented in~\cite{karp98a} and then later revisited in~\cite{exoo08b}, where the case $r \geqslant n/2$ has also been efficiently handled.

\begin{theorem}[\cite{exoo08b}, \cite{karp98a}] \label{Thm_2-fold_lower_bound_refined}
	Let $C \subseteq \F^n$ be an $r$-IdC as well as $s$ and $s'$ be the largest integers such that
	\[
	\sum_{i=1}^{s} \binom{|C|}{i} \leqslant 2^n \ \text{and } \ \sum_{i=0}^{s'} \binom{|C|}{i}  \leqslant 2^n \text.
	\]
\begin{itemize}
\item[(i)] If $r < n/2$, then the cardinality of $C$ satisfies
		\[
		|C| \cdot V(n,r) \geqslant \sum_{i=1}^si\binom{|C|}{i}+(s+1)	\left(2^n-\sum_{i=1}^s\binom{|C|}{i}\right) \text.
		\]
		
		\item[(ii)] If $n/2 \leqslant r \leqslant n-1$, then the cardinality of $C$ satisfies
		\[
		|C| \cdot V(n,n-r-1) \geqslant \sum_{i=1}^{s'}i\binom{|C|}{i}+(s'+1) \left( 2^n-\sum_{i=0}^{s'}\binom{|C|}{i} \right) \text.
		\]
	\end{itemize}
\end{theorem}
Although the previous lower bounds are rather simple, they are still viable when the length $n$ is small compared to the radius $r$ as we will later see in Subsection~\ref{Subsec_Hamming_upper_bounds}. Notice also that the lower bounds do not hold only for the binary Hamming spaces, but they can be generalized for all graphs for which the cardinality of a ball of radius $r$ does not depend on the chosen centre.

\medskip
For the radius $r=1$, the previous lower bounds can be improved by further restricting the number of words covered by exactly two codewords: if $C \subseteq \F^n$ is a $1$-IdC and $\x \in \F^n$ such that $\x$ is covered by exactly $k \ (\geqslant 2)$ codewords, then there exist at most $\binom{k}{2}$ vertices $\y$ satisfying $|I_1(C;\y)| = 2$ and $I_1(C;\y) \subseteq I_1(C;\x)$ (see also Lemma~\ref{Lemma_intersection_balls}). The result was originally presented in~\cite{karp98a} and later a more convenient proof was provided in~\cite{blassBoundsIdentifyingCodes2001}*. Besides a few sporadic lengths $n$, the following lower bound is the best currently known one.

\begin{theorem}[\cite{blassBoundsIdentifyingCodes2001}*, \cite{karp98a}] \label{Thm_father_son_lower_bound}
We have $\Idr{\F^n}{1} \geqslant \dfrac{n2^n}{V(n,2)}$.
\end{theorem}

For the radius $r > 1$, the first efforts to improve the lower bounds of Theorems~\ref{Thm_2-fold_lower_bound} and \ref{Thm_2-fold_lower_bound_refined} were made in~\cite{exoo08}. This approach was further refined in~\cite{exoo08b}.

\subsection{Upper bounds} \label{Subsec_Hamming_upper_bounds}

Concerning the upper bounds, we first focus on the ones related to the radius $r=1$. The exact values of $Id_1(\F^n)$ are known for $n \in \{2, \, 3, \, \ldots, \, 7\}$ as explained in the following.
\begin{itemize}
	\item By Theorem~\ref{Thm_father_son_lower_bound}, we have $Id_1(\F^2) \geqslant 3$, $Id_1(\F^3) \geqslant 4$ and $Id_1(\F^5) \geqslant 10$. Moreover, the lower bounds can be attained by the constructions given in~\cite{karp98a}.
	\item In~\cite{blassBoundsIdentifyingCodes2001}*, the lower bounds $Id_1(\F^4) \geqslant 6$ and $Id_1(\F^7) \geqslant 31$ of Theorem~\ref{Thm_father_son_lower_bound} are respectively improved to $7$ and $32$. Moreover, $1$-IdCs attaining the improved lower bounds are also given.
	\item In~\cite{blassBoundsIdentifyingCodes2001}*, a $1$-IdC in $\F^6$ with $19$ codewords is presented. By Theorem~\ref{Thm_father_son_lower_bound}, it is known that $Id_1(\F^6) \geqslant 18$ and later, in~\cite{exoo08}, it is shown, based on extensive computer searches, that $Id_1(\F^6) = 19$.
\end{itemize}

For $n \geqslant 8$, the smallest cardinalities of $1$-IdCs remain unknown: the best known lower bounds follow by Theorem~\ref{Thm_father_son_lower_bound} and the best constructions have been given in~\cite{charonNewIdentifyingCodes2010}. For an efficient method of constructing 1-IdCs, we require the following definition of $\mu$-fold coverings.
\begin{defin}
	Let $\mu$ be an integer. If $C \subseteq V$ is a code such that $|I_r(C;u)| \geqslant \mu$ for all $u \in V$, then we say that $C$ is a \emph{$\mu$-fold $r$-covering}. Furthermore, if $|I_r(C;u)| = \mu$ for all $u \in V$, then we say that  $C$ is a \emph{perfect} $\mu$-fold $r$-covering.
\end{defin}

New $1$-identifying codes can now be constructed based on 1-IdCs with the additional property that the code is also $2$-fold $1$-covering. The method is based on the classical $\pi(\bu)$-construction discussed, for example, in~\cite[Section~3.4]{cohenCoveringCodes1997}*. For all $\bu \in \F^n$, the \emph{parity-check bit} $\pi(\bu)$ is defined as
\[
\pi(\bu) =
\begin{cases}
	0 & \text{if } w(\bu) \text{ is even}\\
	1 & \text{if } w(\bu) \text{ is odd.}
\end{cases}
\]
\begin{theorem}[\cite{exoo06}] \label{Thm_pi(u)_construction_ell1}
	If $C \subseteq \F^n$ is a $2$-fold $1$-covering and a $1$-IdC in $\F^n$, then
	\[
	C' = \{(\pi(\bu), \bu, \bu + \bc)  \mid \bu \in \F^n, \bc \in C\} \ (\subseteq \F^{2n+1})
	\]
	is a $2$-fold $1$-covering and a $1$-IdC in $\F^{2n+1}$.
\end{theorem}

For the radius $r > 1$, we have the rather simple result from~\cite{blassBinaryCodesIdentification2000}* stating that
\begin{equation} \label{Eq_small_length}
	\Idr{\F^{r+1}}{r} = 2^{r+1}-1 \text.
\end{equation}
For larger radii, other exact values of $\Idr{\F^{n}}{r}$ are rather scarcely known. In what follows, we focus more closely on the exact values for the radii $r=2$ and $r=3$.

\begin{itemize}
\item In~\cite{blassBinaryCodesIdentification2000}*, it has been shown that $\Idr{\F^{4}}{2} = 6$; besides giving a $1$-IdC with 6 codewords, the authors also improve the lower bound of Theorem~\ref{Thm_2-fold_lower_bound_refined} from $5$ to $6$. In the same paper, the constructions with $6$ and $8$ codewords are presented to attain the lower bounds $\Idr{\F^{5}}{2} \geqslant 6$ and $\Idr{\F^{6}}{2} \geqslant 8$ of Theorem~\ref{Thm_2-fold_lower_bound_refined}, respectively.
	
	\item In~\cite{exoo06}, a $2$-IdC in $\F^7$ with $14$ codewords is given. Furthermore, it is shown by an exhaustive computer search that smaller $2$-IdCs in $\F^7$ do not exist.
	
	\item In~\cite{honkalaIdentifyingRadiusCodes1999}*, a $3$-IdC in $\F^5$ with $10$ codewords is given. Furthermore, in~\cite{exoo08b},  the lower bound $\Idr{\F^{5}}{3} \geqslant 9$ of Theorem~\ref{Thm_2-fold_lower_bound_refined} is improved to $10$ by a careful analysis of the proof of the theorem in that particular case.
	
	\item In~\cite{charonNewIdentifyingCodes2010}, $3$-IdCs in $\F^6$ and $\F^7$ with $7$ and $8$ codewords have been presented to attain the lower bounds $\Idr{\F^{6}}{3} \geqslant 7$ and $\Idr{\F^{7}}{3} \geqslant 8$, respectively.
\end{itemize}

For larger radii, the best known lower bounds mostly follow 
by Theorem~\ref{Thm_2-fold_lower_bound_refined} (and its refinement in~\cite{exoo08b}), 
when $n$ is large enough. The best known constructions are mainly due to~\cite{charonNewIdentifyingCodes2010}, although progress has also been made in~\cite{exoo08c} and \cite{exoo08}. In addition, a construction with the asymptotically smallest possible cardinality is presented in the following theorem from~\cite{honk02b}*.

\begin{theorem}[\cite{honk02b}*] \label{Thm_asymptotic_construction_for_ell1}
	Let $0 \leqslant \rho < 1$ and let ${H(x) = -x\log_2{x} - (1-x)\log_2(1-x)}$ denote the binary entropy function. Then: $\displaystyle
	\lim_{n \to \infty} \frac{\log_2{\Idr{\F^n}{\lfloor \rho n \rfloor}}}{n} = 1 - H(\rho)$.
\end{theorem}

\subsection{Conjectures in binary Hamming spaces} \label{Subsec_Hamming_conjectures}

In this subsection, we consider three conjectures concerning identifying codes in binary Hamming spaces. In what follows, we first introduce the following (rather plausible) conjectures.
\begin{itemize}
	\item[(i)] In~\cite{moncelMonotonicityMinimumCardinality2006}, it has been shown that $\Idr{\F^n}{1} \leqslant \Idr{\F^{n+1}}{1}$ for all $n \geqslant 2$. It is also stated as an open question whether the same monotonicity result also holds for larger radii $r > 1$. However, in~\cite{lobsteinLocatingdominationIdentification2020}, it has been noted that, for example, $\Idr{\F^6}{5} = 63$ (by~\eqref{Eq_small_length}) and $31 \leqslant \Idr{\F^7}{5} \leqslant 32$ (by~\cite{charonNewIdentifyingCodes2010}). Hence, the monotonicity does not hold for all $n$, but it still seems plausible that there exists an integer $n(r)$ such that $\Idr{\F^n}{r} \leqslant \Idr{\F^{n+1}}{r}$ for all $n \geqslant n(r)$. However, no progress towards this conjecture has been made.
	
\item[(ii)] In~\cite{blassBoundsIdentifyingCodes2001}* and \cite{karp98a}, it has been conjectured that $\Idr{\F^{n+m}}{r+t} \leqslant \Idr{\F^n}{r}\Idr{\F^m}{t}$. Although there is definite evidence that the conjecture should hold, no proof has been presented; even the significantly simpler statement $\Idr{\F^{n+1}}{1} \leqslant 2\Idr{\F^{n}}{1}$ still remains open.
	
\item[(iii)] In the seminal paper~\cite{karp98a}, the monotonicity of the size of an optimal $r$-IdC in $\F^n$ with respect to the radius $r$ is discussed. In~\cite{exoo08d}, it has been shown that there exists an integer $n_r$ such that $\Idr{\F^n}{r+1} \leqslant \Idr{\F^{n}}{r}$ when $n \geqslant n_r$. However, the requirement for $n_r$ is rather crude and improvements on that seem plausible.
\end{itemize}

In what follows, we concentrate more closely on Conjecture~(ii). We first present some results concerning the weaker formulation $\Idr{\F^{n+1}}{1} \leqslant 2\Idr{\F^{n}}{1}$ of the conjecture. In~\cite{blas99a}* and~\cite{blassBoundsIdentifyingCodes2001}*, first attempts towards this goal have been presented.

\begin{theorem}[\cite{blassBoundsIdentifyingCodes2001}*]
	Let $C$ be a $1$-IdC in $\F^n$.
	\begin{itemize}
		\item[(1)]  The direct sum $C \oplus \F$ is a $1$-IdC in $\F^{n+1}$ if and only if for all codewords $\bc \in C$ there exists another codeword $\bc' \in C$ such that $d(\bc,\bc') = 1$.
		
		\item[(2)] The direct sum $C \oplus \F^2$ is a $1$-IdC in $\F^{n+2}$. In particular, we have $\Idr{\F^{n+2}}{1} \leqslant 4\Idr{\F^{n}}{1}$.
	\end{itemize}
\end{theorem}

In the following theorem (by~\cite{exoo08c}), it is shown that $\Idr{\F^{n+1}}{1} \leqslant (2+\varepsilon_n)\Idr{\F^{n}}{1}$, where $\varepsilon_n \to 0$ as $n \to \infty$.

\begin{theorem}[\cite{exoo08c}]
Assume that $n \geqslant 2$. Then we have $\displaystyle \Idr{\F^{n+1}}{1} \leqslant \left( 2+\frac{1}{n+1} \right) \Idr{\F^{n}}{1}$.
\end{theorem}

Next we focus on the general formulation $\Idr{\F^{n+m}}{r+t} \leqslant \Idr{\F^n}{r}\Idr{\F^m}{t}$ of the conjecture. In~\cite{exoo08c}, various weaker versions of the conjecture have been proved. In particular, it has been shown that $\Idr{\F^{n+m}}{r+1} \leqslant 4\Idr{\F^n}{r}\Idr{\F^m}{1}$. Furthermore, in~\cite{exoo08d}, it has been proved that the conjecture holds when $m$ is relatively small in comparison to $n$.
\begin{theorem}[\cite{exoo08d}]
	Let $r$ and $t$ be integers such that $r \geqslant 2$ and $t \geqslant 2$, and let $n \geqslant 2(t+r)$. If
	\[
	2t\le a \le \frac{2n}{9(t+r)}\left( \frac{2^{r+3}(t-1)!(r-1)!}{9(9(t+r))^{r}}\right)^\frac{1}{t} \textrm{,}
	\]
	then we have
	\[
	\Idr{\F^{n}}{r+t} \leqslant \Idr{\F^{a}}{t}\Idr{\F^{n-a}}{r} \textrm{.}
	\]
\end{theorem}

The following theorem, which has been presented in~\cite{exoo06}, is also closely related to the generalized formulation of Conjecture~(ii).
\begin{theorem}[\cite{exoo06}] \label{Thm_direct_sum_composition}
	Let $r$ and $n_i$ be positive integers. Then we have
	\[
	\Idr{\F^{n_1+n_2+ \cdots + n_r}}{r} \leqslant \prod_{i=1}^{r} \Idr{\F^{n_i}}{1} \text.
	\]
\end{theorem}

\subsection{Identifying sets of vertices} \label{Subsec_larger_ell}

In this subsection, we study $(r,\leqslant \ell)$-IdCs in binary Hamming spaces. 

First efforts on identifying sets of vertices in $\F^n$ were already made in the seminal paper~\cite{karp98a}. In what follows, we first focus on the case with the radius $r=1$. The following theorem states that a $(1, \leqslant \ell)$-IdC is always a $(2\ell-1)$-fold $1$-covering. The theorem was presented in the case $\ell = 2$ in~\cite{honkalaCodesIdentifyingSets2001}* and later generalized to the case $\ell > 2$ in~\cite{laihonenCodesIdentifyingSets2001}. Its proof is based on an insightful application of Lemma~\ref{Lemma_intersection_balls}.

\begin{theorem}[\cite{honkalaCodesIdentifyingSets2001}*, \cite{laihonenCodesIdentifyingSets2001}] \label{Thm_ell_identifying_lower_bound}
	If $C$ is a $(1, \leqslant \ell)$-IdC, then $C$ is also a $(2\ell-1)$-fold $1$-covering. This further implies that
	\begin{equation} \label{Eq_mu_fold_covering_lower_bound}
	\Idrl{\F^n}{1}{\ell} \geqslant \left\lceil (2\ell-1)\frac{2^n}{n+1} \right\rceil \text.
	\end{equation}
\end{theorem}

Inequality~\eqref{Eq_mu_fold_covering_lower_bound} of the previous theorem follows from~\cite[Chapter~14]{cohenCoveringCodes1997}*. The converse of the previous theorem also holds for $\ell \geqslant 3$ as shown in~\cite{laih02a} (based again on Lemma~\ref{Lemma_intersection_balls}).

\begin{theorem}[\cite{laih02a}] \label{Thm_char_ell_identifying}
	Let $\ell$ be an integer such that $\ell \geqslant 3$. A code $C \subseteq \F^n$ is $(1, \leqslant \ell)$-identifying if and only if it is a $(2\ell-1)$-fold $1$-covering.
\end{theorem}

Determining the cardinalities of $\mu$-fold $t$-coverings in $\F^n$ is a well-studied problem. Combining the results of~\cite[Chapter~14]{cohenCoveringCodes1997}* with the previous theorem, the following corollary is obtained.

\begin{corollary}[\cite{laih02a}]
	Let $\ell$ be an integer such that $\ell \geqslant 3$. Then
	$\displaystyle 
	\Idrl{\F^n}{1}{\ell} = (2\ell-1)\frac{2^n}{n+1} \;
	$
if and only if there exist integers $i \geqslant 0$ and $\mu_0 > 0$ such that $\mu_0$ divides $2\ell-1$, and $2\ell-1 \leqslant 2^i\mu_0$, and $n = \mu_02^i-1$.
\end{corollary}

Notice that the requirement $\ell \geqslant 3$ is essential for Theorem~\ref{Thm_char_ell_identifying}. Indeed, in~\cite{rantoTwoFamiliesOptimal2002}*, it has been shown that there exists an infinite family of $3$-fold $1$-covering codes which are not $(1,\leqslant 2)$-identifying. However, in the same paper, optimal constructions for $(1,\leqslant 2)$-IdCs have been presented based on the so-called $\pi(\bu)$-construction already applied in Theorem~\ref{Thm_pi(u)_construction_ell1}.

\begin{theorem}[\cite{rantoTwoFamiliesOptimal2002}*]
	If $C \subseteq \F^n$ is a $(1, \leqslant 2)$-IdC, then
	\[
	C' = \{(\pi(\bu), \bu, \bu + \bc)  \mid \bu \in \F^n, \bc \in C\} \ (\subseteq \F^{2n+1}) \text,
	\]
	where $\pi(\bu)$ denotes the parity-check bit of $\bu$, is a $(1, \leqslant 2)$-IdC in $\F^{2n+1}$. This further
implies that $\Idrl{\F^{2n+1}}{1}{2} \leqslant 2^n\Idrl{\F^{n}}{1}{2}$.
\end{theorem}

In~\cite{rantoTwoFamiliesOptimal2002}*, it has been proved that there exist $(1, \leqslant 2)$-IdCs $C_1 \subseteq \F^5$ and $C_2 \subseteq \F^7$ with $16$ and $48$ codewords, respectively. Hence, the codes $C_1$ and $C_2$ attain the lower bound of Theorem~\ref{Thm_ell_identifying_lower_bound} implying that $C_1$ and $C_2$ are also perfect $3$-fold $1$-coverings. Applying the previous theorem to the codes $C_1$ and $C_2$, we obtain two infinite families of $(1, \leqslant 2)$-IdCs which are perfect $3$-fold $1$-coverings attaining the lower bound of Theorem~\ref{Thm_ell_identifying_lower_bound}. Hence, an optimal $(1, \leqslant 2)$-IdC is obtained for lengths $n \geqslant 5$ for which a perfect $3$-fold $1$-covering also exists (see ~\cite[Theorem~14.2.4]{cohenCoveringCodes1997}*). More precisely, an optimal $(1, \leqslant 2)$-IdC in $\F^n$ is obtained for the lengths $n = 3 \cdot 2^k - 1$ and $n = 2^{k+2} - 1$, where $k$ is a positive integer.

To find $(1, \leqslant \ell)$-IdCs for the lengths other than the ones with perfect $(2\ell-1)$-fold $1$-coverings, the direct sum construction of the following theorem may be used; for $\ell = 2$, the result is due to~\cite{honkalaCodesIdentifyingSets2001}*, and for $\ell > 2$, the result is presented in~\cite{laih02a}.

\begin{theorem}[\cite{honkalaCodesIdentifyingSets2001}*, \cite{laih02a}] \label{Thm_direct_sum_construction_ell_r1}
	Let $\ell$ be an integer such that $\ell \geqslant 2$. If $C \subseteq \F^n$ is a $(1, \leqslant \ell)$-IdC, then so is the direct sum $C \oplus \F \subseteq \F^{n+1}$.	In particular, we have
	$\displaystyle 
	\Idrl{\F^{n+1}}{1}{\ell} \leqslant 2\Idrl{\F^{n}}{1}{\ell} \text.
	$
\end{theorem}

Note that the theorem proves, for $\ell \geqslant 2$, the issue similar to the conjecture ${\Idr{\F^{n+1}}{1} \leqslant 2\Idr{\F^{n}}{1}}$ mentioned in Subsection~\ref{Subsec_Hamming_conjectures}.
The generalized conjecture $\Idr{\F^{n+m}}{r+t} \leqslant \Idr{\F^n}{r}\Idr{\F^m}{t}$ can also be approached for $\ell \geqslant 2$. In fact, the conjecture has been partially solved in the case $\ell \geqslant r+3$ in~\cite{exoo08c}.

\begin{theorem}[\cite{exoo08c}]
If $\ell \geqslant r+3$, then
	$\displaystyle 
	\Idrl{\F^{n+m}}{r+1}{\ell} \leqslant \Idrl{\F^n}{1}{\ell}\Idrl{\F^m}{r}{\ell} \text.$
\end{theorem}

Next, we consider lower bounds on $(r, \leqslant \ell)$-IdCs when the radius $r \geqslant 2$. Firstly, a lower bound similar to the one of Theorem~\ref{Thm_2-fold_lower_bound_refined} has been provided in~\cite[Theorem~3.4]{exoo08b}. It gives the best known lower bounds on $(r, \leqslant \ell)$-IdCs when $r \geqslant 2$ and $n$ is relatively small compared to $r$ and $\ell$. For larger lengths $n$, the best known lower bounds are given by the following theorem due to~\cite{exoo08b} and \cite{exoo08}.

\begin{theorem}[\cite{exoo08b}, \cite{exoo08}]
	For $r \geqslant 2$ and $\ell \geqslant 2$, we have
	\[
	\Idrl{\F^{n}}{r}{\ell} \geqslant \left\lceil \frac{(2\ell-2)2^n}{\binom{n}{r}} \right\rceil \ \text{ and } \ \Idrl{\F^{n}}{r}{\ell} \geqslant \left\lceil \frac{(2\ell-1)2^n}{\binom{n}{r}+\binom{n}{r-1}}  \right\rceil \text.
	\]
\end{theorem}

In what follows, we continue with the constructions of $(r, \leqslant \ell)$-IdCs when the radius $r \geqslant 2$. As discussed earlier, the cardinalities $\Idrl{\F^{n}}{r}{\ell}$ are rather well-known for $r=1$ and $\ell \geqslant 2$. Hence, the approach of the following theorem by~\cite{exoo08}, which generalizes Theorem~\ref{Thm_direct_sum_composition}, is quite effective for constructing $(r, \leqslant \ell)$-IdCs.

\begin{theorem}[\cite{exoo08}] \label{Thm_direct_sum_composition_larger_ell}
	For $r \geqslant 1$ and $\ell \geqslant 2$, we have
	$\displaystyle 
	\Idrl{\F^{n_1+n_2+ \cdots + n_r}}{r}{\ell} \leqslant \prod_{i=1}^{r} \Idrl{\F^{n_i}}{1}{\ell} \text.
	$
\end{theorem}

Another approach for constructing $(r, \leqslant \ell)$-IdCs is based on a generalization of the direct sum construction of Theorem~\ref{Thm_direct_sum_construction_ell_r1}. This approach is studied in the following theorem, which has been presented in~\cite{exoo08b}.

\begin{theorem}[\cite{exoo08b}]  \label{Thm_direct_sum_construction_ell_r}
	Let $\ell$ be an integer such that $\ell \geqslant 2$.
	\begin{itemize}
		\item[(1)] If $1 \leqslant r \leqslant 2$ and $C$ is an $(r, \leqslant \ell)$-IdC in $\F^n$, then $C \oplus \F^r$ is an $(r, \leqslant \ell)$-IdC in $\F^{n+r}$. In particular, we have $\Idrl{\F^{n+r}}{r}{\ell} \leqslant 2^r\Idrl{\F^{n}}{r}{\ell}$.
		\item[(2)] If $r \geqslant 3$ and $C$ is an $(r, \leqslant \ell)$-IdC in $\F^n$, then $C \oplus \F^{r+1}$ is an $(r, \leqslant \ell)$-IdC in $\F^{n+r+1}$. In particular, we have $\Idrl{\F^{n+r+1}}{r}{\ell} \leqslant 2^{r+1}\Idrl{\F^{n}}{r}{\ell}$.
	\end{itemize}
\end{theorem}

Most of the best known $(r, \leqslant \ell)$-IdCs are due to Theorems~\ref{Thm_direct_sum_composition_larger_ell}~and~\ref{Thm_direct_sum_construction_ell_r}. However, some incremental improvements have been made in~\cite{dhanalakshmiConstructionsRidentifyingCodes2019}. In addition, an asymptotic result similar to Theorem~\ref{Thm_asymptotic_construction_for_ell1} for $\ell = 1$ has been proved 
 in~\cite{jans09}.

\begin{theorem}[\cite{jans09}]
	Let $\ell \geqslant 1$ be fixed and let $\rho \in [0,1/2)$. If $r/n \to \rho$ as $n \to \infty$, then
	\[
	\lim_{n \to \infty} \frac{\log_2{\Idrl{\F^n}{r}{\ell}}}{n} = 1 - H(\rho) \text,
	\]
	where $H(x) = -x\log_2{x} - (1-x)\log_2(1-x)$ is the binary entropy function.
\end{theorem}

\subsection{Other variants} \label{other_variants}

Finally, we briefly discuss some variants of $(r,\leqslant \ell)$-IdCs. We first consider a situation
where, with respect to the illustration proposed in Section~\ref{IdLD6Intro}, a malfunctioning processor may incorrectly report whether a problem occurs in its ball. Formally, that leads to the following definition (by~\cite{honk02d}*), which is in view of Section~\ref{IdLD6Conc} presented for general graphs $G=(V,E)$.
\begin{defin}[\cite{honk02d}*] \label{Def_strongly}
	Let $C \subseteq V$ be a code in $G=(V,E)$, and $r$ and $\ell$ be nonnegative integers. For $X \subseteq V$, we define
	$\displaystyle 
	\mathcal{I}_r(X) = \{ U \mid I_r(X) \setminus (X \cap C) \subseteq U \subseteq I_r(X)\}
	$. 
	If for all distinct $X_1, X_2 \subseteq V$ with $|X_1| \leqslant \ell$ and $|X_2| \leqslant \ell$ we have $\mathcal{I}_r(X_1) \cap \mathcal{I}_r(X_2) \neq \emptyset$, then we say that $C$ is a \emph{strongly $(r, \leqslant \ell)$-IdC}.
\end{defin}

For strongly $(1, \leqslant \ell)$-IdCs in binary Hamming spaces, similar results as the ones described in the case of (usual) $(1, \leqslant \ell)$-IdCs have been derived in~\cite{honk02d}*,~\cite{laih02c}~and~\cite{laih02b}. Their proofs are also based on a careful analysis of the $\mu$-fold $1$-covering properties of the codes. Besides strongly identifying codes, various robust identifying codes, which are able to withstand additions and/or deletions of edges in the underlying graph, have been considered for binary Hamming spaces in~\cite{honk06d}*,~\cite{honkalaVertexrobustIdentifyingCodes2010}*,~\cite{laihonenOptimalEdgerobustVertexrobust2005}~and~\cite{laihonenEdgerobustLeqIdentifying2007}.

Recall that $(r, \leqslant \ell)$-IdCs can locate up to $\ell$ faulty processors in the considered network. However, if more than $\ell$ faulty processors are attempted to be located using an $(r, \leqslant \ell)$-IdC, then errors may occur and we might even be unaware that the set of faulty processors indicated by the code is incorrect. This issue has been addressed in~\cite{honkalaNewClassIdentifying2007}*, where a new class of so-called \emph{$(r, \leqslant \ell)^+$-IdCs} has been introduced. These codes can \emph{locate} up to $\ell$ faulty processors and \emph{detect} if more than $\ell$ faulty processors exist in the network. In particular, for the binary Hamming spaces $\F^n$, it has been shown in~\cite{honkalaNewClassIdentifying2007}* that $C \subseteq \F^n$ is a $(1, \leqslant \ell)^+$-IdC if and only if $C$ is a $(2 \ell + 1)$-fold $1$-covering.\vspace*{-2mm}

\section{Infinite grids}\label{IdLD6Specif4}

\begin{figure}[h!]
\vspace*{1mm}
\begin{center}
\includegraphics*[scale=0.75]{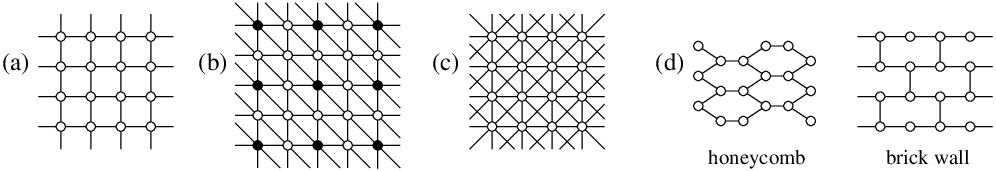}
\end{center}\vspace*{-5mm}
\caption{Partial representations of the four grids: (a)~the square grid $G_S$; (b)~the triangular grid $G_T$: black vertices are codewords (cf. Theorem~\ref{6triangleMark6}); (c)~the king grid $G_K$; (d)~the hexagonal grid $G_H$ (with two possible representations: as a honeycomb or as a brick wall).}\label{fig6grilles6}\vspace*{-2mm}
\end{figure}

We survey now results related to infinite grids.
The four infinite $2$-dimensional grids $G_S$ (square), $G_T$ (triangular), $G_K$ (king), and $G_H$ (hexa\-gonal) have $\mathbb{Z} \times \mathbb{Z}$ as their vertex sets. They are partially represented in Figure~\ref{fig6grilles6} and are formally defined in the next subsections.

 \begin{figure}[b!]
 \vspace*{-2mm}
  \begin{center}
\includegraphics*[scale=0.75]{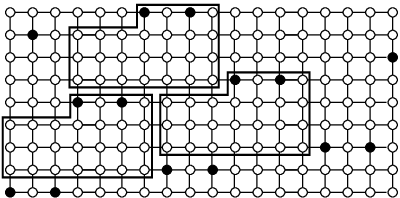}
\end{center}\vspace*{-5mm}
\caption{A periodic $5$-IdC in the square grid $G_S$, of density $2/25$; codewords are in black.}
\label{fig6SqGr66}
\end{figure}

Because these grids are infinite graphs, we consider here the {\it density} of a code $C$ instead of the cardinality of $C$. Let $G$ denote one of these grids. For any integer $q$, let $Q_q$ denote the set of vertices $(x, y) \in \mathbb{Z} \times \mathbb{Z}$ with $|x| \leqslant q$ and $|y| \leqslant q$. The density of a code $C$ in $G$ is defined by $\partial(C,G)=\limsup _{q \rightarrow \infty}\frac{|C\, \cap \, Q_q|}{|Q_q|}$. Then we define the \textit{minimum density} $\partial^{Id}_{r}(G)$ of $G$ as the minimum of $\partial(C, G)$ over the set of the $r$-IdCs $C$ of $G$. Thus, the problem considered in this section consists in computing the values of $\partial^{Id}_{r}(G)$ for $r \geqslant 1$ and for $G$ belonging to $\{G_S, G_T, G_K, G_H\}$.

In the following subsections, we give lower and upper bounds for $\partial^{Id}_{r}(G)$, for $r \geqslant 1$ and $G \in \{G_S$,
$G_T, G_K, G_H\}$. Several constructions of codes are obtained by heuristics searching for small subcodes inside tiles that will be repeated periodically, as done in~\cite{char02a} (see Figure~\ref{fig6SqGr66} for such an example dealing with the square grid and $r=5$). These constructions provide upper bounds for $\partial^{Id}_{r}(G)$.

\subsection{The square grid} \label{IdLD6Specif6411} 
\begin{figure} [h!]
  \begin{center}
\includegraphics*[scale=0.75]{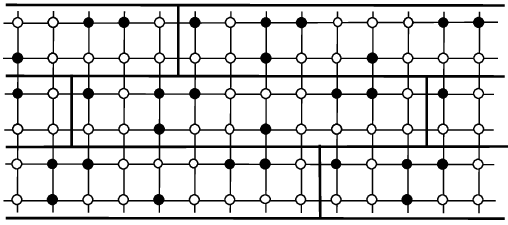}       
\end{center}\vspace*{-3mm}
\caption{A periodic $1$-IdC in the square grid $G_S$, of density $7/20$; codewords are in black.}
\label{ADDfig01}\vspace*{-1mm}
\end{figure}

The square grid, $G_S$, has vertex set $V_S=\mathbb{Z} \times \mathbb{Z}$ and edge set $E_S=\{\{u, v\}: u-v\in \{(1,0),(0,1)\}\}$. The exact minimum density is known for $r=1$, as stated by Theorem~\ref{6SqGIDr16}. The value 7/20 as an upper bound for $\partial^{Id}_{1}(G_{S})$ comes from~\cite{cohe99a}* and from~\cite{lits04} as a lower bound. Figure~\ref{ADDfig01} displays a construction providing the upper bound.

\begin{theorem}[\cite{lits04}, \cite{cohe99a}*] \label{6SqGIDr16} For $r = 1$, $\partial^{Id}_{1}(G_{S})=\frac{7}{20}$. 
\end{theorem}

For $r \geqslant 2$, Theorem~\ref{6squareme46} specifies bounds for $\partial^{Id}_{r}(G_{S})$. The lower bound for $r=2$ comes from~\cite{junn13}; the general lower bound for $r \geqslant 3$ comes from~\cite{char01a}*; all the upper bounds come from~\cite{honk02a}*.

\begin{theorem} \label{6squareme46}

(a) For $r=2$, $\frac{6}{35}\approx 0.17143 \leqslant \partial^{Id}_{2}(G_{S})\leqslant \frac{5}{29}\approx 0.17241$.

(b) For $r\geqslant 3$, $\frac{3}{8r+4} \leqslant \partial^{Id}_{r}(G_{S})\leqslant \left\{ \begin{array}{ll}
    \frac{2}{5r} & :r\mbox{ even }\\
    \frac{2r}{5r^2-2r+1} & :r\mbox{ odd.}
  \end{array}\right.$
\end{theorem}

When $r$ increases, these bounds are close to $\frac{3}{8r}=\frac{0.375}{r}$ and $\frac{2}{5r} = \frac{0.4}{r}$. The previous upper bounds have been improved in~\cite{char02a}, using heuristics, for $r\in \{3,4,5,6\}$ (with respectively ${1/8 = 0.125}$, ${8/85 \approx 0.094}$, ${2/25 = 0.08}$ and ${3/46 \approx 0.065}$).

\subsection{The triangular grid} \label{IdLD6Specif6412} 
The triangular grid, or square grid with one diagonal, $G_T$, has vertex set $V_T=\mathbb{Z} \times \mathbb{Z}$ and edge set $E_T=\{\{u, v\}: u-v\in \{(1,0),(0,1),(1,-1)\}\}$.

As for the square grid, the exact minimum density is known for $r=1$:

\begin{theorem} [\cite{karp98a}]\label{6triangleMark6}  
   For $r = 1$, $\partial^{Id}_{1}(G_{T})=\frac{1}{4}$.
\end{theorem}

This minimum value is provided by the construction displayed in Figure~\ref{fig6grilles6}(b), which has the property that every codeword is dominated by exactly one codeword (itself) and every non-codeword is dominated by exactly two codewords (either horizontally or vertically or diagonally). This proves that the code is indeed $1$-identifying, and moreover that it is best possible. Anyway, it is shown in~\cite{cohe01a}* that this property cannot stand for $r>1$. We have instead the following results for the triangular grid.

\begin{theorem} \label{6triangleus46}
(a) {\rm(\cite{char01a}*)} For $r\geqslant 2$, $\frac{2}{6r+3} \leqslant \partial^{Id}_{r}(G_{T})\leqslant \left\{ \begin{array}{ll}
    \frac{1}{2r+2} & :r\in \{1,2,3\}\;{\rm mod}\;4\\
    \frac{1}{2r+4} & :r=0\;{\rm mod}\;4.
  \end{array}\right.$\vspace*{1mm}

(b) {\rm(\cite{char02a})} We have $\partial^{Id}_{3}(G_{T})\leqslant \frac{2}{17}\approx 0.11765$ and $\partial^{Id}_{5}(G_{T})\leqslant \frac{1}{13}\approx 0.07692$. 
\end{theorem}

\subsection{The king grid} \label{IdLD6Specif6413}
The king grid, or square grid with two diagonals, $G_K$, has vertex set $V_K=\mathbb{Z} \times \mathbb{Z}$ and edge set $E_K=\{\{u, v\}: u-v\in \{(1,0),(0,1),(1,-1),(1,1)\}\}$. Its name comes from the fact that, on an infinite chessboard, the $r$-ball of a vertex~$v$ is the set of vertices that a king, starting from~$v$, can reach in at most $r$ moves.

\medskip
It is remarkable that the best density is known for all $r\geqslant 1$, thanks to a sharp analysis of the neighbourhoods of the vertices. In the following theorem, the value 2/9 as the lower bound on $\partial^{Id}_{1}(G_{K})$ comes from~\cite{cohe01b}* and from~\cite{char02a} as the upper bound.

\begin{theorem} \label{6kingglory46} (a) {\rm(\cite{char02a}, \cite{cohe01b}*)}  For $r=1$, $\partial^{Id}_{1}(G_{K})= \frac{2}{9}\approx 0.22222$.\vspace*{1mm}

(b) {\rm(\cite{char03a}*)} For $r>1$, $\partial^{Id}_{r}(G_{K})= \frac{1}{4r}$.
\end{theorem}

\begin{figure} [h!]
\vspace{2mm}
  \begin{center}
\includegraphics*[scale=0.75]{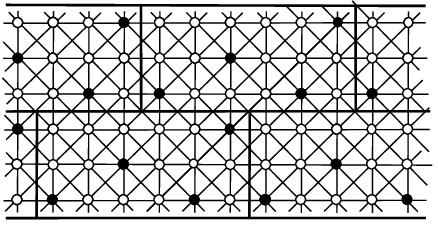}
\hspace{1cm}
\includegraphics*[scale=0.75]{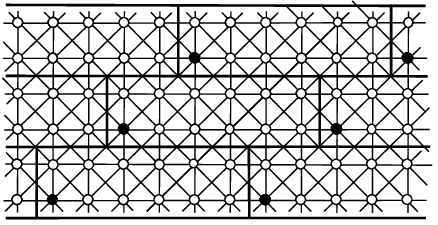}      
\end{center}\vspace*{-3mm}
\caption{A periodic $1$-IdC (left) and a periodic 3-IdC (right) in the king grid $G_K$, of density 2/9 and 1/12 respectively; codewords are in black.}
\label{Fig_King}
\end{figure}

Figure~\ref{Fig_King} displays a 1-IdC with minimum density 2/9 on the left and a 3-IdC with minimum density 1/12 on the right. In fact, for $r \geqslant 2$, the periodic pattern of the $r$-IdC is similar to the one displayed here for $r=3$: a rectangle with two rows and $2r$ columns, with only one codeword, located in the bottom left corner; the patterns are concatenated horizontally and then the resulting strip is replicated vertically with a shift of two columns. It is shown in~\cite{char03a}* that this construction is optimal.

\subsection{The hexagonal grid} \label{IdLD6Specif6414}  
The hexagonal grid, $G_H$, has vertex set $V_H=\mathbb{Z} \times \mathbb{Z}$ and edge set $E_H=\{\{u, v\}: u=(i,j)$ and $u-v\in \{(0,(-1)^{i+j+1}),(1,0)\}\}$. It is the grid with the sparsest and weakest results. 

For $r=1$, the following upper bound is from~\cite{salo_2024}, improving the one provided by~\cite{cohe00a}* (equal to $3/7 \approx 0.42857$), the lower bound is from~\cite{cuki13}.
\begin{theorem}[\cite{cuki13}, \cite{salo_2024}] \label{6hexaGIDr16} For $r = 1$, $\frac{5}{12}\approx 0.41667 \leqslant \partial^{Id}_{1}(G_{H})\leqslant \frac{53}{126} \approx 0.42063$. 
\end{theorem} 


In addition, it should be mentioned that an improvement $23/55 \approx 0.41818$ to the lower bound has been claimed in the  manuscript~\cite{stolee_2014}.

The minimum density is known for $r=2$, as specified by the next theorem. In this one, the lower bounds in (a) and (b) come from \cite{junn12b} and~\cite{mart10}, respectively, and both upper bounds from~\cite{char02a}; the general lower bounds in~(c) and (d) come from~\cite{char01a}*, and the upper bounds from~\cite{stan11b}.

\begin{theorem}[\cite{char01a}*, \cite{char02a}, \cite{junn12b}, \cite{mart10}, \cite{stan11b}] \label{6hexmultiple6}
  (a) For $r = 2$, $\partial^{Id}_{2}(G_{H})=\frac{4}{19}\approx 0.21053$.

  (b) For $r = 3$, $\frac{3}{25}=0.12 \leqslant \partial^{Id}_{3}(G_{H}) \leqslant \frac{1}{6}\approx 0.16667$.

  (c) For $r$ even with $r \geqslant 4$, $\frac{2}{5r+3} \leqslant \partial^{Id}_{r}(G_{H})\leqslant \frac{5r+3}{6r(r+1)}$.

  (d) For $r$ odd with $r \geqslant 5$, $\frac{2}{5r+2} \leqslant \partial^{Id}_{r}(G_{H})\leqslant \frac{5r^2+10r-3}{(6r-2)(r+1)^2}$.

\end{theorem}

When $r$ increases, these bounds are close to $\frac{2}{5r}$ and $\frac{5}{6r}\approx \frac{0.83333}{r}$. Better upper bounds, close to $1/(r+4)$ for $r$ even or to $1/(r+5)$ for $r$ odd, are known for several values of $r$ with $r \leqslant 30$, obtained in \cite{char02a} by the use of heuristics.





\section{Structural properties of twin-free graphs}
\label{IdLD6SI1}
We shall now review some works related to the study of structural problems. First (Subsection~\ref{IdLD6RWOParam65}), we consider some classic parameters of graphs, and we investigate the following question: what are the extremal values that these parameters can achieve when we restrict ourselves to $r$-twin-free graphs? In Subsection~\ref{IdLD6SI12}, we study the question when it is possible to remove a vertex from an $r$-twin-free graph while keeping the property of being $r$-twin-free, leading to the notion of \textit{terminal graph}. Then, we study the effects of removing a vertex (Subsection~\ref{IdLD6SI32}) or an edge (Subsection~\ref{IdLD6SI3}) on the $r$-Id numbers.


For $S \subset V$, we denote by~$G \setminus S$ the graph obtained from~$G$ by deleting the vertices of~$S$. Similarly, for an edge~$e$ of $G$, we note by $G \setminus \{e\}$ the graph obtained from $G$ by removing~$e$.

\subsection{Extremal values of usual parameters in twin-free graphs}
\label{IdLD6RWOParam65}

In this subsection, we investigate the extremal values that different usual parameters of graphs can achieve in a connected $r$-twin-free graph. In accordance with \cite{auge09b}*, we consider the following para\-meters: the size~$\varepsilon$ (i.e., the number of edges), the minimum degree~$\delta$, the radius~$\rho$, the diameter~$D$, the size~$\alpha$ of a maximum independent set (see~\cite{lobsteinLocatingdominationIdentification2020} for references dealing with other parameters).

More precisely, if $\pi$ stands for such a parameter, we search for the smallest value, $f_\pi(r, n)$, that $\pi$ can reach over the set of connected and $r$-twin-free graphs with exactly $n$ vertices ($n \geqslant 2r+1$): \\[1mm]
\centerline{$f_\pi(r, n)=\min\{\pi(G): G$ connected, $r$-twin-free with $n$ vertices$\}$.} \\[1mm]
 We also consider the largest value that $\pi$ can reach, leading to the quantity $F_\pi(r, n)$: \\[1mm]
\centerline{$F_\pi(r, n)=\max\{\pi(G): G$ connected, $r$-twin-free with $n$ vertices$\}$.}\\

\noindent {\bf $\bullet$ Number of edges $\varepsilon$}\\
For any connected graph with $n$ vertices, the size is at least $n-1$. The path $P_n$ on $n$ vertices meets this bound. Thus, for any $r \geqslant 1$ and $n \geqslant 2r + 1$, $f_\varepsilon(r,n)=n-1$.

The maximum number of edges possible for $r$-twin-free graphs with $n$ vertices is known for $r=1$, and can be achieved only by the complete graph $K_n$ minus a maximum matching~\cite{dani03}.

\begin{theorem} \label{6edgesJM65} {\rm(\cite{dani03})} For $r = 1$ and $n\geqslant 3$, $F_\varepsilon(1,n)=\frac{n(n-1)}{2} - \lfloor \frac{n}{2}\rfloor$. 
\end{theorem}

For $r>1$, we are close to the exact value.

\begin{theorem} \label{6edgesUSr65} {\rm(\cite{auge09b}*)} (a) For $r = 2$ and $n$ large enough, $\frac{n^2}{2}-2n\log_2n \lesssim F_\varepsilon(2,n) \lesssim \frac{n^2}{2}-\frac{1}{2}n\log_2n$.

  (b) For $r>2$ and $n$ large enough with respect to $r$, \\
  \centerline{$\frac{n^2}{2}-rn\log_2n \lesssim F_\varepsilon(r,n) \lesssim \frac{n^2}{2}-0.63(r-0.915)n\log_2n$.}
\end{theorem}
\medskip


\noindent {\bf $\bullet$ Minimum degree $\delta$}\\ 
The path $P_n$ provides the value of $f_\delta(r,n)$ for any $r$ and any $n\geqslant 2r+1$: the minimum of $\delta$ for a connected and $r$-twin-free graph is equal to 1. Case~(b) below relies on the fact that the complete graph minus a maximum matching is $1$-twin-free (cf. Theorem~\ref{6edgesJM65}). On the other hand, the minimum degree cannot be too large with respect to the ratio $n/r$, otherwise there will be $r$-twins; Theorem~\ref{6mindelta1USr65} specifies bounds stated thanks to this ratio.

\begin{theorem} \label{6mindelta1USr65} {\rm(\cite{auge09b}*)} (a) For $r\geqslant 1$ and $n \geqslant 2r+1$, $f_\delta(r,n)=1$.

  (b) For $r=1$ and $n\geqslant 3$, $F_\delta(1,n)=n-2$.

  (c) For $r=2$ and $n\geqslant 5$, $F_\delta(2,n)=\lfloor \frac{n-2}{2}\rfloor$.

  (d) For $r\geqslant 3$, $F_\delta(r,2r+1)=1$.

  (e) For $r\geqslant 3$ and $n\geqslant 2r+2$, set $k=\lfloor \frac{n-2}{r}\rfloor$. Then:\vspace*{1mm}

  \indent \indent (e$_1$) $k-1 \leqslant F_\delta(r,n)$ if $k$ is odd, and $k \leqslant F_\delta(r,n)$ if $k$ is even;

  \indent \indent (e$_2$) for $3 \leqslant r \leqslant 5$, $F_\delta(r,n)\leqslant \frac{n}{\lfloor \frac{r}{2}\rfloor +1}-1$;

   \indent \indent (e$_3$) for $r\geqslant 6$, $F_\delta(r,n)\leqslant \min \left\{ \frac{n}{\lfloor \frac{r}{2}\rfloor +1}-1, \frac{3n-r+2}{2(r-5)}\right\}$. 
\end{theorem}

It is conjectured in \cite{auge09b}* that $F_\delta(r,n)$ is equal to $\lfloor \frac{n-2}{r}\rfloor$ for any $r \geqslant 1$ and any $n \geqslant 2r + 1$.\\

\noindent {\bf $\bullet$ Radius $\rho$}\\
In any connected graph of order~$n$, the radius is between $1$ (the complete graph, the star) and $\lfloor \frac{n}{2}\rfloor$ (the path, the cycle). The study of the radius in $r$-twin-free graphs is easy, the results are complete.
\begin{theorem} \label{6minradius2USr65} {\rm(\cite{auge09b}*)} (a) For $r\geqslant 1$ and $n\geqslant 2r+1$, $f_\rho(r,n)=r$.

  (b) For $r\geqslant 1$ and $n\geqslant 2r+1$, $F_\rho(r,n)=\lfloor \frac{n}{2}\rfloor$. 
\end{theorem}

\begin{figure} [h!]
\vspace*{2mm}
  \begin{center}
\includegraphics*[scale=1.22]{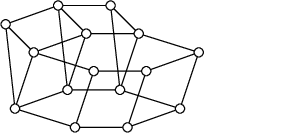}
\end{center}\vspace*{-3mm}
\caption{The case $n=15$, $r=3$ of Theorem~\ref{6mindiamm2USr65}, Case (b); this graph is $3$-twin-free and has diameter~${r+1=4}$.}
\label{figIddiamm6SI}\vspace*{-5mm}
\end{figure}

\noindent {\bf $\bullet$ Diameter $D$}\\
The results on the diameter in $r$-twin-free graphs are also complete. Once again, Case (a) and Case (c) are related to the path $P_{2r+1}$, which is the only connected and $r$-twin-free graph on $2r+1$ vertices, and more generally to the path $P_n$. Figure~\ref{figIddiamm6SI} illustrates Case~(b).

\begin{theorem} \label{6mindiamm2USr65} {\rm(\cite{auge09b}*)} (a) For $r\geqslant 1$, $f_D(r, 2r+1)=2r$.

  (b) For $r\geqslant 1$ and $n\geqslant 2r+2$, $f_D(r, n)=r+1$.

  (c) For $r\geqslant 1$ and $n\geqslant 2r+1$, $F_D(r, n)=n-1$. 
\end{theorem}

\noindent {\bf $\bullet$ Size $\alpha$ of a maximum independent set}\\
In any connected graph of order~$n$ which is not the complete graph, $\alpha$ lies between $2$ and $n-1$ (the star). Theorem~\ref{6easyId61} above, from~\cite{Auger_2008}, contributes to Case~(b) and to the lower bound of Case (c) in the following theorem.

\begin{theorem} \label{6minalpha2USr65} {\rm(\cite{auge09b}*)} (a) For $r=1$ and $n\geqslant 3$, $f_\alpha(1,n)=2$.

  (b)  For $r\geqslant 2$, $f_\alpha(r,2r+1)=r+1.$

  (c) For $r\geqslant 2$ and $n\geqslant 2r+2$, set $k=\lfloor \frac{n-2}{r}\rfloor$. Then: $r+1\leqslant f_\alpha(r,n) \leqslant \left\{ \begin{array}{ll}
     \frac{2n}{k+2} \text{ for } k \mbox{ even}\\
     \frac{2n}{k+1} \text{ for } k \mbox{ odd.}
  \end{array}\right.$ 
\end{theorem}

The star gives Case~(a) in Theorem~\ref{6minalpha3USr65}.

\begin{theorem} \label{6minalpha3USr65} {\rm(\cite{auge09b}*)} (a) For $r = 1$ and $n\geqslant 3$, $F_\alpha(1,n)=n-1$.

  (b) For $r\geqslant 2$ and $n\geqslant 2r+1$, let $k$ be the largest integer with $k+r\lceil \log_2k\rceil \leqslant n-1$. Then: $\max\left\{\lceil \frac{n}{2}\rceil, k+\lceil \log_2k\rceil \lfloor \frac{r}{2}\rfloor\right\}\leqslant F_\alpha(r,n) \leqslant n-r$. 
\end{theorem} 

If $n$ is large with respect to $r$, then $k+\lceil \log_2k\rceil \lfloor \frac{r}{2}\rfloor$ behaves approximately like $n-\frac{r}{2}\log_2n$, and so behaves also the lower bound of $F_\alpha(r,n)$ in Case (b).\\

\subsection{Terminal graphs}
\label{IdLD6SI12}
We need a new definition in this section: the one of an \textit{r-terminal graph}.

\begin{defin}
 If $G=(V,E)$ is $r$-twin-free, then we say that $G$ is $r$-{\it terminal} if for all $v\in V$, $G \setminus \{v\}$ is not $r$-twin-free.
 \end{defin}

 In other words, $G$ is not $r$-terminal if there exists a vertex $v\in V$ such that $G \setminus \{v\}$ is also $r$-twin-free. We are interested in the structure and number of $r$-terminal graphs, if they exist. 

Observe that, for $r > 1$, $P_{2r+1}$ (which is 1-twin-free, see Theorem~\ref{6easyId61}) is $r$-terminal: indeed, removing any vertex from $P_{2r+1}$ leaves a graph with $r$-twins. For $r=1$, the case of $P_3$ is particular: removing the middle vertex yields two isolated vertices, constituting a $1$-twin-free graph. This leads us to address the following questions:
\begin{itemize}
\item (a)~Are there 1-terminal graphs?
\item (b)~For $r>1$, are there $r$-terminal graphs other than $P_{2r+1}$? If so, how many are there?
\end{itemize}

As detailed below, the answer to~(a) is negative. The answer to~(b) is multifold: it is negative if we restrict ourselves to trees. But it is positive for $r\geqslant 3$ otherwise; the case $r=2$ remains open.\\

\noindent {\bf $\bullet$ The case $r=1$}

There is no 1-terminal graph. Indeed, as observed above, for $n=3$, $P_3$ is not 1-terminal. For $n \geqslant 4$, consider a vertex $v$ such that $V \setminus \{v\}$ is a 1-IdC of $G$ (by Theorem~\ref{lemmeNath6}, such a vertex does exist). Then we can show that the graph $G \setminus \{v\}$, which may be connected or not, is still 1-twin-free and, hence, $G$ is not 1-terminal.

The following theorem sharpens this result with respect to connectivity.
\begin{theorem}
\label{theoIdLD6SI12b} {\rm(\cite{char06}*)} Let $n\geqslant 4$ and $G=(V,E)$ be any connected 1-twin-free graph of order~$n$. Then there exists $v\in V$ such that $G \setminus \{v\}$ is 1-twin-free and connected. 
\end{theorem}

\noindent {\bf $\bullet$ The case of trees}

For $r = 1$, the previous result shows that there is no 1-terminal tree. For $r > 1$, $P_{2r+1}$ is the only $r$-terminal tree, as specified by the next theorem.

\begin{theorem}
\label{theoIdLD6SI12c} {\rm(\cite{char06}*)} Let $r\geqslant 1$, $n\geqslant 2r+2$ and $T=(V,E)$ be any $r$-twin-free tree of order~$n$. Then there exists a leaf $v \in V$ such that $T \setminus \{v\}$ is $r$-twin-free (and connected). Consequently, the only $r$-terminal tree for $r>1$ is the path $P_{2r+1}$. 
\end{theorem}

\begin{figure} [h!]
  \begin{center}
\includegraphics*[scale=0.5]{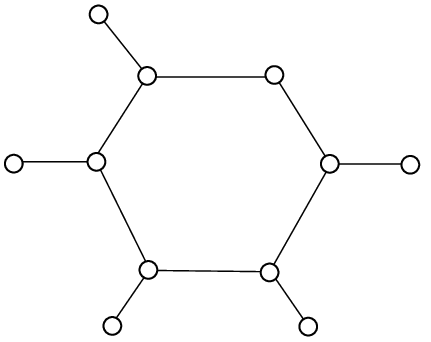}
\end{center}\vspace*{-4mm}
\caption{A 3-terminal graph.}
\label{G_3-term} \vspace*{-3mm}
\end{figure}

\noindent {\bf $\bullet$ The case $r\geqslant 3$}\\
We now come back to general graphs. Then, for $r \geqslant 3$, $P_{2r+1}$ is not the only $r$-terminal graph. One possible construction in order to obtain another $r$-terminal graph is the following: take a cycle of length $2r$ and add a neighbour to each vertex of the cycle but one; Figure~\ref{G_3-term} illustrates this construction for $r=3$. Theorem~\ref{G-term} is more specific:

\begin{theorem}
\label{G-term} {\rm(\cite{char06}*)} For $r\geqslant 3$, $P_{2r+1}$  is not the only $r$-terminal graph. For $r\geqslant 6$, there are infinitely many $r$-terminal graphs.
\end{theorem}

Observe that the construction of Theorem~\ref{G-term} does not work for $r=2$; the problem remains open:

\begin{op_pb}
Apart from $P_5$, do $2$-terminal graphs exist?
\end{op_pb}

Another open problem is the situation for $r\in \{3,4,5\}$:

\begin{op_pb}
For $r\in \{3,4,5\}$, is the number of $r$-terminal graphs finite or infinite?
\end{op_pb}


\subsection{Removing a vertex}
\label{IdLD6SI32}
\begin{figure} [h!]
  \begin{center}
\includegraphics*[scale=0.75]{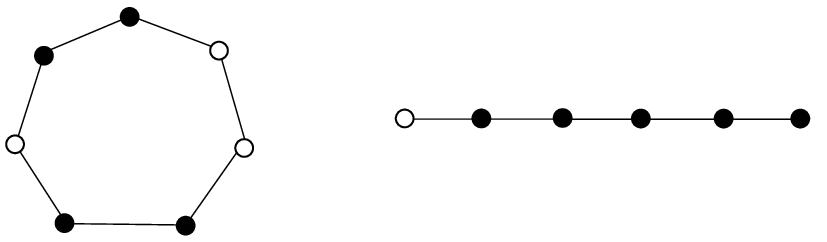}
\end{center}\vspace*{-3mm}
\caption{Optimal 2-IdCs for $C_7$ and $P_6$; codewords are in black.}
\label{Rob08_fig}
\end{figure}

The problem considered now is the following: given an $r$-twin-free graph $G=(V,E)$ and a vertex $v\in V$, and assuming that the graph $G \setminus \{v\}$ is $r$-twin-free, what can be said about $Id_{r}(G \setminus \{v\})$ with respect to $Id_{r}(G)$? It is easy to observe that removing a vertex may increase $Id_{r}(G)$. For instance, removing a vertex from the cycle $C_7$ on 7 vertices increases $Id_2$ by 1: $Id_2(C_7) = 4$ and $Id_2(P_6) = 5$ (see Figure~\ref{Rob08_fig}).

\noindent {\bf $\bullet$ The case $r=1$}\\
It is known from~\cite{fouc11} that, if $G$ and $G \setminus S$ are 1-twin-free, then we have $Id_1(G) - Id_1(G \setminus S) \leqslant |S|$. In particular,
for $v \in V$, $Id_1(G) - Id_1(G \setminus \{v\}) \leqslant 1$; moreover, the two inequalities are tight. Thus, removing one vertex cannot imply a large decrease of $Id_1$: at most 1.

\medskip
But removing a vertex may imply a large increase of $Id_1$, as shown in~\cite{char12b}*.

\begin{figure} [h!]
\vspace*{1mm}
  \begin{center}
\includegraphics*[scale=0.75]{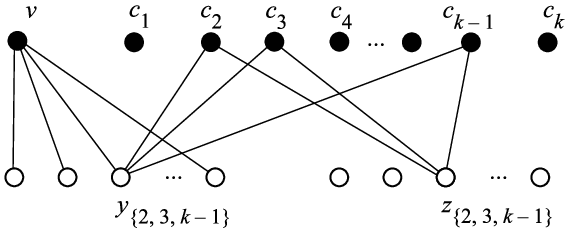}
\end{center}\vspace*{-4mm}
\caption{A partial representation (for $I = \{2, 3, k-1\}$) of the graphs involved in the proof of Theorem~\ref{Th_1_vertex_less}; codewords are in black.}\label{Fig_1_vertex_less}
\end{figure}

\begin{theorem} [\cite{char12b}*] \label{Th_1_vertex_less}
There exist (connected) graphs $G$ with $n$ vertices and a vertex $v$ of $G$ with $Id_1(G \setminus \{v\}) - Id_1(G) \approx 0.5n - 1.5 \log_2 n$ and $Id_1(G \setminus \{v\}) / Id_1(G) \approx 0.5n/\log_2 n$.
\end{theorem}
The graphs involved in this theorem are built in the following way (see Figure~\ref{Fig_1_vertex_less} for a partial representation of them). They are connected and bipartite. We start with $k$ vertices $c_i$ ($1 \leqslant i \leqslant k$), which will be, as well as the vertex $v$, the codewords of $G$. Then, for each subset $I \subseteq \{1, 2, ..., k\}$ of indices with at least two elements ($|I| \geqslant 2$), we create two vertices $y_I$ and $z_I$: $y_I$ is linked to $v$ and to all the vertices $c_i$ with $i \in I$, while $z_I$ is linked to all the vertices $c_i$ with $i \in I$ but not to $v$. Thus, we obtain a graph $G$ with $n = 2(2^k - k - 1) + k + 1 = 2^{k+1} - k - 1$ vertices. As the subsets $I$ are not singletons, $C = \{v, c_1, c_2, ..., c_k\}$ is a 1-IdC of $G$, with about $\log_2 n$ codewords. But, if we remove $v$, then $y_I$ and $z_I$ become almost twins for every $I$: the only way to separate them is to select one of them as a codeword, which, finally, requires to select about at least $n/2$ codewords.

M. Pelto provided a sharper result in \cite{pelt13} (with graphs which are not necessarily connected):
\begin{theorem}
\label{theoIdLD6SI321b}
{\rm(\cite{pelt13})} Let $n$ be the order of~$G$, $S$ a proper subset of $V$ and $v \in V$.

(a) If $n\geqslant 2^{|S|-1}$, then $Id_1(G \setminus S) - Id_1(G)\leqslant n-2|S|- \left \lfloor \dfrac{n-|S|}{2^{|S|}} \right \rfloor $; the inequality is tight for $n$ sufficiently large with respect to~$|S|$.

(b) For $|S| = 1$: $Id_1(G \setminus \{v\}) - Id_1(G) \leqslant \left \lfloor\dfrac{n}{2} \right \rfloor -\varepsilon$, with $\varepsilon =2$ for $n\in \{2,4,5,6,8\}$ and $\varepsilon =1$ otherwise; the inequality is tight.

(c) If $G$ is bipartite, then  $Id_1(G \setminus \{v\}) - Id_1(G)\leqslant \left \lfloor \dfrac{n-\log_2(n-\log_2n)}{2} \right \rfloor - 1$; the inequality is tight. 
\end{theorem}

So, for $r=1$, if one vertex is deleted, the identification number cannot drop by more than one, and can increase by a quantity close to $n/2$.\\

\noindent  {\bf $\bullet$ The case $r\geqslant 2$}\\
Now both $Id_{r}(G \setminus \{v\}) - Id_{r}(G)$ and $Id_{r}(G) - Id_{r}(G \setminus \{v\})$ can be large: ${Id_{r}(G \setminus \{v\}) - Id_{r}(G)}$ can increase to approximately $\frac{n}{4}$ (for even~$r$) and $\frac{n(3r-1)}{12r}$ (for odd~$r$), or $\frac{n(2r-2)}{2r+1}$, according to whether we want the graphs to be connected or not, and $Id_{r}(G) - Id_{r}(G \setminus \{v\})$ to approximately $\frac{n(r-1)}{r}$.

\begin{theorem}
\label{theoIdLD6SI322a}
 {\rm(\cite{char12b}*)}
(a) Let $r\geqslant 2$. There exist connected graphs $G$ of order $n$ and a vertex $v$ of $G$ with ${Id_{r}(G \setminus \{v\}) - Id_{r}(G) \geqslant \dfrac{(n-1)(2r-2)}{2r+1}-2r}$.

(b) Let $r\geqslant 2$ be even. There exist connected graphs $G$ of order $n$ and a vertex $v$ of $G$ such that $G \setminus \{v\}$ is connected and ${Id_{r}(G \setminus \{v\}) - Id_{r}(G) \geqslant \dfrac{n}{4}-(r+1)}$.

(c) Let $r\geqslant 3$ be odd. There exist connected graphs $G$ of order $n$ and a vertex $v$ of $G$ such that $G \setminus \{v\}$ is connected and ${Id_{r}(G \setminus \{v\}) - Id_{r}(G) \geqslant \dfrac{n(3r-1)}{12r}-r}$.

\end{theorem}

\begin{theorem}
\label{theoIdLD6SI322b}
 {\rm(\cite{char12b}*)}
 There exist connected graphs $G$ of order $n$ with $n = 2$ mod $r$ and a vertex $v$ of $G$ such that $G \setminus \{v\}$ is connected and $Id_{r}(G) - Id_{r}(G \setminus \{v\})\geqslant \dfrac{(n-3r-1)(r-1)+1}{r}$. 
\end{theorem}

\begin{figure}[h!]
\begin{center}
\includegraphics*[scale=0.6]{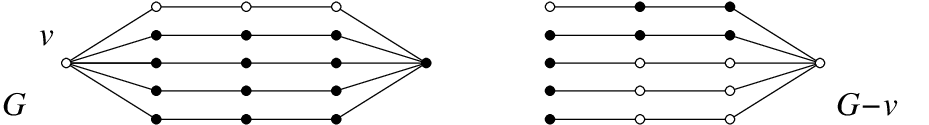}
\end{center}\vspace*{-3mm}
\caption{Illustration of Theorem~\ref{theoIdLD6SI322b} for $r=3$: in both graphs, the black vertices form an optimal $3$-IdC; ${Id_{3}(G) - Id_{3}(G \setminus \{v\})=13-8=5}$.}\label{figIdLD6SI} \vspace*{-5mm}
\end{figure}

The proof of this theorem is based on graphs obtained by a collection of paths $P_{r}$ plus two extra vertices, including $v$, linked to the two extremities of each path; Figure~\ref{figIdLD6SI} illustrates Theorem~\ref{theoIdLD6SI322b} for $n=17$, $r=3$.

\subsection{Removing an edge}
\label{IdLD6SI3}
Let $G$ be a graph and $e$ an edge of $G$.
The problem considered here is similar to the previous one, but with the deletion of an edge instead of a vertex: given an $r$-twin-free graph $G=(V,E)$ and an edge $e \in E$, and assuming that the graph $G \setminus \{e\}$ is $r$-twin-free, what can be said about $Id_{r}(G \setminus \{e\})$ with respect to $Id_{r}(G)$? \\

\noindent {\bf $\bullet$ The case $r=1$}

\begin{theorem}[\cite{char12a}*]
  \label{theoIdLD6SI311a} 
If $G$ and $G \setminus \{e\}$ are 1-twin-free, then $Id_1(G) - Id_1(G \setminus \{e\}) \in \{-2,-1,0,1,2\}$ and these values can be reached by pairs of connected graphs~$G$ and $G \setminus \{e\}$.
\end{theorem}

\noindent {\bf $\bullet$ The case $r\geqslant 2$}\\
Now the differences $Id_{r}(G) - Id_{r}(G \setminus \{e\})$ and $Id_{r}(G \setminus \{e\}) - Id_{r}(G)$
can be large, and we obtain results which slightly vary with~$r$.

\begin{theorem}[\cite{char12a}*]
  \label{theoIdLD6SI312a} 
Let $k\geqslant 2$ be an arbitrary integer.

\medskip
(a1) Let $r\geqslant 2$. There exist a graph $G$ with ${(r+1)k+r\lceil \log_2(k+2)\rceil +2r}$ vertices and an edge~$e$ of $G$ with $Id_{r}(G)\geqslant k$ and $Id_{r}(G \setminus \{e\}) \leqslant r\lceil \log_2(k+2)\rceil +r +3$.

(a2) Let $r\geqslant 5$. There exist a graph $G$ with ${(2r-2)k+r\lceil \log_2(k+2)\rceil +r +3}$ vertices and an edge~$e$ of $G$ with $Id_{r}(G)\geqslant k$ and $Id_{r}(G \setminus \{e\}) \leqslant r\lceil \log_2(k+2)\rceil +r +1$.

(b1) For $r=2$, there exist a graph $G$ with ${3k+2\lceil \log_2(k+2)\rceil +5}$ vertices and an edge~$e$ of $G$ with ${Id_{2}(G \setminus \{e\})\geqslant k}$ and $Id_{2}(G) \leqslant 2\lceil \log_2(k+2)\rceil +5$.

(b2) Let $r\geqslant 3$. There exist a graph $G$ with ${(2r-2)k+r\lceil \log_2(k+2)\rceil +r +2}$ vertices and an edge~$e$ of $G$ with $Id_{r}(G \setminus \{e\})\geqslant k$ and $Id_{r}(G) \leqslant r\lceil \log_2(k+2)\rceil +r +1$. 
\end{theorem}
Closely related results can be obtained if we require that $G \setminus \{e\}$ is connected.

\medskip
Note that $k$ can be chosen arbitrarily and is linked to the order~$n$ of~$G$ and~$G \setminus \{e\}$ by the relation $n=(c_1r+c_2)k+r\lceil \log_2(k+2)\rceil +(c_3r+c_4)$ where the integer quadruple ($c_1,c_2,c_3,c_4)$ takes different values in (a1), (a2), (b1) and~(b2) above. This means, roughly speaking, that $k$ is a fraction, depending on~$r$, of~$n$; therefore, Theorem~\ref{theoIdLD6SI312a} implies that, given $r\geqslant 2$, there is an infinite collection of graphs~$G$ of order~$n$ and two positive constants $\alpha$ and $\beta$ with $Id_{r}(G)\geqslant \alpha n$ and, after deletion of a suitable edge~$e$, $Id_{r}(G \setminus \{e\})\leqslant \beta \log_2 n$ (or the other way round: $Id_{r}(G)\leqslant \beta \log_2 n$ and $Id_{r}(G \setminus \{e\})\geqslant \alpha n$). We can see that adding or deleting one edge can lead to quite a drastic difference for the identification numbers.

\medskip
We studied above what the identification numbers can become when adding or deleting vertices or edges. But we can also be interested in what an {\it existing} (optimal) code becomes when edges or vertices are deleted or added. This leads to the notion of {\it robustness}: an $r$-Id code $C$ is $t$-{\it edge-robust}~\cite{honk06f}* in $G$ if $C$ remains an $r$-IdC in all the graphs obtained from $G$ by adding or deleting edges, with a total amount of additions and deletions at most~$t$. This issue is dealt with, among others, in~\cite{honk06d}* and in~\cite{laih06a}. Different definitions exist for $t$-vertex-robust codes, see~\cite{honk06f}* or~\cite{ray03a}.

\section{Number of optimal codes}
\label{IdLD6NumbSol}
In this section, we are interested in graphs with many optimal codes. In his PhD thesis \cite{pelt12} under Iiro Honkala's supervision, M. Pelto worked on the notion of completely different optimal identifying codes in infinite grids. During his PhD defense the following question arose: how large can the number of optimal $r$-IdCs be? Considering once again our example of the location of a faulty processor in a multiprocessor network, this means that we want not only to use the smallest possible number of controllers,
but also to have a large number of choices for their locations.

Let $\nu_r(n)$ be the maximum number of optimal (labelled) $r$-IdCs that a graph on $n$ vertices may have. It is easy to observe that $\nu_r(n)$ can be exponential with respect to $n$. Indeed, consider a cycle $C_{2r+2}$ on $2r+2$ vertices. Any set of $2r + 1$ vertices is an optimal $r$-IdC of $C_{2r+2}$ and there are $2r + 2$ such sets. Replicating $C_{2r+2}$ $k$ times provides a (disconnected) graph on ${k(2r+2)}$ vertices and with $(2r + 2)^k$ optimal $r$-IdCs. Hence the lower bound $(2r + 2)^k \leqslant \nu_r(k(2r+2))$, i.e., when $n$ is divisible by $2r+2$, ${(2r + 2)^{n/(2r+2)} = 2^{n \log_2(2r+2)/(2r+2)} \leqslant \nu_r(n)}$. In particular, for $r=1$, we obtain the lower bound $2^{n/2} \leqslant \nu_1(n)$ when $n$ is divisible by 4.

\medskip
In order to obtain a larger number of optimal 1-IdCs, the notion of \textit{separating-only code} is defined in \cite{honk14c}* as follows for $r=1$:

\begin{defin}\label{def_SOC}
Let $G=(V, E)$ be a graph. A subset $C$ of $V$ is said to be a separating-only code (SOC) of $G$ if:
\begin{itemize}
\item $\exists \, v_0 \in V: I_1(C;v_0) = \emptyset$;
\item $\forall \, v_1 \in V, \, \forall \, v_2 \in V$ with $v_1 \neq v_2: \; I_1(C;v_1) \neq I_1(C;v_2).$
\end{itemize}
\end{defin}

Let $\sigma(G)$ denote the size of a minimum SOC of $G$ and let $\mu(G)$ denote the number of optimal SOCs of $G$.\\

\noindent \textbf{Remarks.} \\
1. Some 1-twin-free graphs do not admit any SOC (for instance, the path on three vertices).\\
2. In a SOC $C$, the vertex $v_0$ with $I_1(C;v_0) = \emptyset$ in Definition~\ref{def_SOC} is unique (with respect to $C$).\\
3. Since adding the vertex $v_0$ of Definition~\ref{def_SOC} to the associated SOC leads to a 1-IdC, we have the inequalities $Id_1(G) - 1 \leqslant \sigma(G) \leqslant Id_1(G)$.\\

\begin{figure} [h!]
\vspace*{1mm}
  \begin{center}
\includegraphics*[scale=1.8]{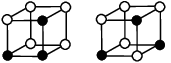}
\end{center}\vspace*{-5mm}
\caption{The two non-isomorphic optimal SOCs of $\mathbb{F}^3$; the three black vertices are the codewords.}
\label{fig_SOC}
\end{figure}

The following discussion is based on~\cite{honk14c}*. Consider the binary Hamming space $\mathbb{F}^3$, with 8 vertices. Figure~\ref{fig_SOC} displays the two non-isomorphic unlabelled optimal SOCs of $\mathbb{F}^3$, each requiring three codewords: $\sigma(\mathbb{F}^3) = 3$. There are 24 (respectively 8) labelled SOCs isomorphic to the SOC on the left (respectively right) of Figure~\ref{fig_SOC}: $\mu(\mathbb{F}^3) = 32$.

\begin{figure} [h!]
\vspace*{1mm}
  \begin{center}
\includegraphics*[scale=1.8]{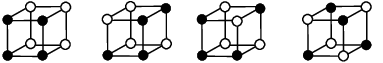}
\end{center}\vspace*{-5mm}
\caption{The four non-isomorphic optimal 1-IdCs of $\mathbb{F}^3$; the black vertices are the codewords.}
\label{fig_SOC_2}
\end{figure}

On the other hand, Figure~\ref{fig_SOC_2} displays the four non-isomorphic unlabelled optimal 1-IdCs of $\mathbb{F}^3$: ${Id_1(\mathbb{F}^3) = 4}$. There are 6 labelled 1-IdCs isomorphic to the first unlabelled model, 24 to the second, 24 to the third and 2 to the fourth: $\nu_1(\mathbb{F}^3) = 56$.

\begin{figure} [h!]
  \begin{center}
\includegraphics*[scale=1.8]{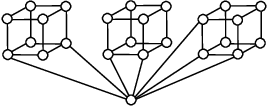}
\end{center}\vspace*{-3mm}
\caption{The graph $G_1$; only some edges are represented between the vertices of the three copies of $\mathbb{F}^3$ and the extra
vertex at the bottom.} \label{G1}
\end{figure}

Now, consider three copies of $\mathbb{F}^3$ and add an extra vertex linked to the 24 vertices of these three copies (see Figure~\ref{G1}). We obtain the first graph, $G_1$, of a series of graphs $G_q$ described below. In order to construct optimal 1-IdCs of $G_1$, we consider the following two constructions:
\begin{itemize}
\item combine optimal 1-IdCs of each $\mathbb{F}^3$;
\item or select the extra vertex, choose one copy of $\mathbb{F}^3$ and select an optimal SOC in it, then combine them with optimal 1-IdCs inside the other two copies of $\mathbb{F}^3$.
\end{itemize}

\noindent In both cases, all the vertices, including the extra vertex, are covered and separated from the others. More precisely, we obtain optimal 1-IdCs of $G_1$. Observe that $G_1$ has 25 vertices with $\nu_1(G_1) = 56 \times 56 \times 56 + 3 \times 32 \times 56 \times 56 = 476\,672$. Replicating $G_1$ $k$ times provides a (disconnected) graph with $n = 25k$ vertices and $476\,672^{k} = 2^{n\log_2(476\,672)/25} \approx 2^{0.7545n}$ optimal 1-IdCs.

\begin{figure} [h!]
  \begin{center}
\includegraphics*[scale=2]{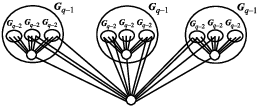}
\end{center}\vspace*{-3mm}
\caption{The scheme of the graph $G_q$ ($q \geqslant 3$).}
\label{Gq}
\end{figure}

\medskip
In fact, this result is improved in \cite{honk14c}* by designing a series of connected graphs $G_q$ defined in a similar way as $G_1$: for $q \geqslant 2$, $G_{q}$ is obtained from $G_{q-1}$ by replicating $G_{q-1}$ three times and by adding an extra vertex linked to the vertices of the three copies of $G_{q-1}$ (see Figure~\ref{Gq}).

Finally, based on the graphs $G_q$, we obtain the following theorem:

\begin{theorem} \label{6bcpoptimr=167}  {\rm(\cite{honk14c}*)} For an infinite number of integers $n$, there exist connected graphs of order $n$ admitting approximately $2^{0.770 n}$ different optimal 1-IdCs. 
\end{theorem}
For $r>1$, other constructions, based on trees admitting many optimal $r$-IdCs, are proposed, leading to the next theorem, where $\frac{1+\log_2 5}{2}$ is approximately equal to 0.664.

\begin{theorem} \label{6bcpoptimtoutr67}  {\rm(\cite{honk14c}*)} Let $r\geqslant 1$ be an integer and $\varepsilon >0$ be a real. For an infinite number of integers $n$, there exist connected graphs of order $n$ admitting $2^{\left(\frac{1+\log_2 5}{2}-\varepsilon\right)n}$ different optimal $r$-IdCs. 
\end{theorem}

Note that a study of the set of the optimal $r$-identifying codes in twin-free graphs can also be found in~\cite{HHL_2016}*.

\section{Variants and related concepts}
\label{IdLD6Conc}
We have tried to give an overview of several contributions by Iiro Honkala on identifying codes. Some variants could have been considered too.

\medskip
For instance, with respect to the context exposed in Section 1, we could discuss the situation when a processor cannot control itself, i.e., when we consider the open neighbourhood ${N_r(v) = B_r(v) \setminus \{v\}}$ instead of the whole ball $B_r(v)$ for any vertex $v$, leading to the concept of {\it open neighbourhood} identification. This was introduced in~\cite{honk02d}* (see also~\cite{seo10}). This definition is one of several definitions for so-called {\it fault-tolerant codes}, where different scenarii are considered for the alarms given by the codewords.

The strongly $(r, \leqslant \ell)$-IdCs (of Definition~\ref{Def_strongly}) are also related to open neighbourhoods. Indeed, for $\ell = 1$, the definition can be reformulated as follows: a code $C$ is strongly $(r, \leqslant 1)$-identifying if for all $v_1\in V$, $v_2\in V$ with $v_1\neq v_2$, the sets $\{B_{r}(v_1)\cap C, N_{r}(v_1)\cap C\}$ and $\{B_{r}(v_2)\cap C, N_{r}(v_2)\cap C\}$ are disjoint. See~\cite{honk10}*, where the best density for a strongly $(1, \leqslant 1)$-IdC in the triangular grid is proved to be $6/19\approx 0.3158$, or \cite{laih02c} and~\cite{laih02b}.


A more drastic variant consists in adopting other patterns instead of the $r$-balls, as in \cite{honk03a}* (see also, e.g.,~\cite{fouc13c}, \cite{junn11} or \cite{rose03a}), especially for the grids defined in $\mathbb{Z} \times \mathbb{Z}$ where the patterns may be no longer isotropic (for example, we may adopt a horizontal path centred -- or not... -- on $v$ instead of $B_r(v)$).

But surely the most studied variant of $r$-IdCs is the one of $r$-\textit{locating-dominating codes} (see~\cite{lobsteinLocatingdominationIdentification2020}). They are defined as follows:

\begin{defin}
An $r$-{\it locating-dominating code}~$C$ of a graph $G = (V, E)$ is a subset of $V$ fulfilling the two properties:
\begin{itemize}
\item $\forall \, v \in V: I_r(C;v) \neq \emptyset$;
\item $\forall \, v_1 \in V \setminus C, \, \forall \, v_2 \in V \setminus C \text{ with } v_1 \neq v_2: \; I_r(C;v_1) \neq I_r(C;v_2).$
\end{itemize}
\end{defin}

So, with respect to $r$-IdCs, the difference is that the $r$-separation property must be fulfilled only by non-codewords (thus, any $r$-IdC is also an $r$-locating-dominating code). Locating-dominating codes were introduced by P.~J. Slater in 1983~\cite{Slater_83} (for a more easily accessible source, see~\cite{Rall_84}). Since this date, they have been extensively studied (see~\cite{lobsteinLocatingdominationIdentification2020} and, for updated references, the ongoing bibliography~\cite{http}), and Iiro Honkala contributed much to their study. But summarizing Iiro's contributions to that topic would require another article...\\

\noindent \textbf{Acknowledgements.}
 The author Ville Junnila was funded in part by the Academy of Finland grant 338797.
The authors thank the referees for their remarks and suggestions that helped to correct some errors and improve the article.

\bibliographystyle{fundam}

\end{document}